\documentclass[aps,prx,superscriptaddress,floatfix,reprint,longbibliography]{revtex4-2}
\usepackage{amsmath}
\usepackage{amsfonts}
\usepackage{amssymb}
\usepackage{setspace}
\usepackage{color}
\usepackage{xcolor}
\usepackage{textcomp}
\usepackage{graphicx}
\usepackage{float}
\usepackage{longtable}
\usepackage{verbatim}
\usepackage[colorlinks=true, linkcolor=blue, citecolor=blue, urlcolor=blue]{hyperref}
\usepackage{upgreek}

\begin{document}

\title{Cavity-Mediated Electron-Electron Interactions:\\Renormalizing Dirac States in Graphene}

\author{Hang Liu}
\email{hang.liu@mpsd.mpg.de}
\affiliation{Max Planck Institute for the Structure and Dynamics of Matter and Center for Free-Electron Laser Science, Luruper Chaussee 149, 22761, Hamburg, Germany}

\author{Francesco Troisi}
\affiliation{Max Planck Institute for the Structure and Dynamics of Matter and Center for Free-Electron Laser Science, Luruper Chaussee 149, 22761, Hamburg, Germany}

\author{Hannes Hübener}
\affiliation{Max Planck Institute for the Structure and Dynamics of Matter and Center for Free-Electron Laser Science, Luruper Chaussee 149, 22761, Hamburg, Germany}

\author{Simone Latini}
\email{simola@dtu.dk}
\affiliation{Department of Physics, Technical University of Denmark, 2800 Kgs. Lyngby, Denmark}
\affiliation{Max Planck Institute for the Structure and Dynamics of Matter and Center for Free-Electron Laser Science, Luruper Chaussee 149, 22761, Hamburg, Germany}

\author{Angel Rubio}
\email{angel.rubio@mpsd.mpg.de}
\affiliation{Max Planck Institute for the Structure and Dynamics of Matter and Center for Free-Electron Laser Science, Luruper Chaussee 149, 22761, Hamburg, Germany}
\affiliation{Initiative for Computational Catalysis, The Flatiron Institute, 162 Fifth Avenue, New York, NY 10010, United States}

\begin{abstract}
Embedding materials in optical cavities has emerged as a strategy for tuning material properties. Accurate simulations of electrons in materials interacting with quantum photon fluctuations of a cavity are crucial for understanding and predicting cavity-induced phenomena. In this article, we develop a non-perturbative quantum electrodynamical approach based on a photon-free self-consistent Hartree-Fock framework to model the coupling between electrons and cavity photons in crystalline materials. We apply this theoretical approach to investigate graphene coupled to the vacuum field fluctuations of cavity photon modes with different types of polarizations.  The cavity photons introduce nonlocal electron-electron interactions, originating from the quantum nature of light, that lead to significant renormalization of the Dirac bands. In contrast to the case of graphene coupled to a classical circularly polarized light field, where a topological Dirac gap emerges, the nonlocal interactions induced by a quantum \textit{linearly} polarized photon mode give rise to the formation of flat bands and the opening of a topologically trivial Dirac gap. When two symmetric cavity photon modes are introduced, Dirac cones remain gapless, but a Fermi velocity renormalization yet indicates the relevant role of nonlocal interactions. These effects 
disappear in the classical limit for coherent photon modes. This new self-consistent theoretical framework paves the way for the simulation of non-perturbative quantum effects in strongly coupled light-matter systems, and allows for a more comprehensive discovery of novel cavity-induced quantum phenomena.   
\end{abstract}

\maketitle

\section{Introduction}
Engineering electromagnetic vacuum field fluctuations via optical cavities is emerging as a novel way to control the ground state of condensed matter systems~\cite{Ebbesen_Science_review, Bloch2022_Nature, Franco_Nori_review, RevModPhys.91.025005, Hertzog_review, Ebbesen_review2016,  Hannes_NM_comment,  Ruggenthaler2018}. Cavity photon vacuum fluctuations have been experimentally demonstrated to control the reaction pathway of molecules by changing the potential energy landscape \cite{Ebbesen_Science}, affect the magneto-transport property of electron gas  for tunable Landau levels and quantum Hall effects \cite{Kono2018,Faist2019,cavity_QHE_exp}, and modify the electronic conductivity of organic and inorganic extended solids \cite{Ebbesen2015,Nature_1T-TaS2_Fausti}. These observations indicate that the vacuum fluctuations in optical cavities can be strong enough to mediate interactions and alter properties of quantum materials. 

To understand and predict the impact of cavity vacuum fluctuations on materials, various theoretical approaches based on Quantum Electrodynamics (QED) have been employed, with the main challenge being the modeling of cavity-mediated interactions, arising from the strong collective coupling of photons with electrons and nuclei.
Photon-mediated interactions of a nonlocal type, beyond perturbation theory,
are crucial for controlling atomic structures \cite{Feist_PRX2015}, charge transfer \cite{PhysRevX.10.041043}, chemical reactivity \cite{Hertzog_review}, and van der Waals forces \cite{molecule_vdw} in molecules, as well as electron topology in atomic chains \cite{1D_chain_Christian, ED_finite_SSH, Daniel_PRB, PhysRevLett.125.217402}. These phenomena can only be grasped by approaches like Hartree-Fock (HF) approximation \cite{ab_HF_molecule}, coupled-cluster expansion \cite{PhysRevX.10.041043}, and exact diagonalization with partial or full configurations \cite{molecule_CI}, which all share the capability of capturing photon-mediated nonlocal electron-electron interactions in a non-perturbative fashion. 
Moreover, quantum electrodynamical density functional theory (QEDFT) within the local-density approximation (LDA) \cite{QED_functional_Christian, ab-initio_Perspective_Dominik} provides another framework for exploring photon-mediated 
interactions and their consequent phenomena. 
It has been used for finite interacting systems, demonstrating cavity-induced donor-acceptor charge transfer \cite{cavity_molecule_PNAS_Angel} and spin glass behavior \cite{Dominik_spin_glass} in molecular clusters.
The QEDFT is non-perturbative by construction, however its functionals have not yet included nonlocal interactions. The present work can serve a guide to develop and improve QEDFT functionals in the future.

For extended crystalline systems, theoretical studies rely on a local approximation of cavity induced interactions, and they are built on perturbative approximations (except for those using QEDFT) \cite{Hannes_review_MQT,review_Sentef}. 
Within perturbation theory, for example, the possibility of cavity-engineered topological phase transitions has been proposed in Su-Schrieffer-Heeger chains \cite{1D_chain_Olesia}, graphene \cite{graphene_Russian, graphene_Sentef, graphene_Vasil, graphene_cavity_topology_single_particle}, and twisted layered materials \cite{cavity_twisted_BG, cavity_twisted_TMD}. Also, modifications of superconducting properties in crystals have been explored using both perturbation theories \cite{PhysRevLett.122.133602,PhysRevLett.122.167002,Sentef_superconductivity,PhysRevLett.127.177002} and non-perturbative QEDFT simulations \cite{I-Te_PNAS}. In the latter case, a local density approximation is employed for the exchange-correlation functional and therefore, if photon-mediated nonlocal interactions are to be included, a different approach has to be devised. 

In this article, we develop a theoretical approach that addresses the coupling between optical cavities and extended crystals in a non-perturbative manner, enabling a precise description of photon-mediated nonlocal (long-range) electron interactions that arise from the quantum nature of cavity photons and have no counterparts for classical light fields. By making a high-frequency expansion of the exact QED Hamiltonian, we arrive at a photon-free Hamiltonian which features both local and nonlocal cavity-mediated electron interactions. It is important to stress that the nonlocal electron-electron interactions are \textit{not} mediated by the Coulomb potential, but they result from the quantum vacuum fluctuations of the cavity photons.
To solve the photon-free Hamiltonian, we perform the HF approximation on this photon-mediated interaction, arriving at what we call the \textit{photon-free QED-HF} formulation, which is notably distinct from the HF method for Coulomb interactions. The direct and exchange interactions in the QED-HF formulation describe a novel kind of nonlocal electron-electron interaction, induced by fluctuating cavity photons, which should not be confused with that in the usual HF theory for the Coulomb electron-electron interactions. The photon-mediated nonlocal interactions arise even in a system of non-interacting electrons, where the coupling between electrons is mediated by the transverse part of the cavity vacuum field fluctuations.

\begin{figure}[htbp]
    \centering
    \includegraphics[width=1.0\linewidth]{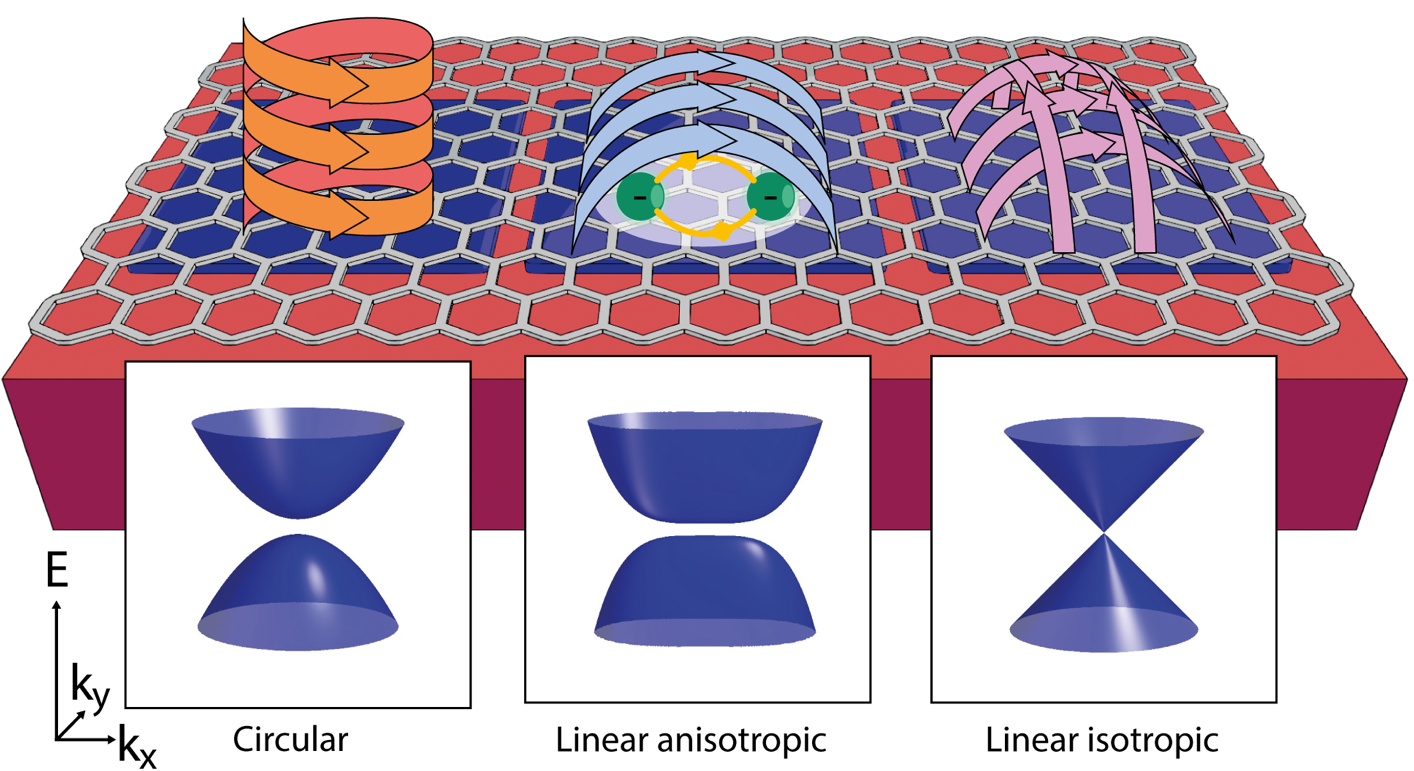}
    \caption{Illustration of the renormalized Dirac cones of monolayer graphene coupled to cavity photon modes of different polarizations. Due to cavity-mediated electron interactions, a circularly polarized photon induces an isotropic Dirac gap with nontrivial band topology, while a linearly polarized photon induces a flat and anisotropic Dirac gap with trivial topology.
    In contrast, two isotropic linearly polarized photons, with the same frequency and amplitude and the perpendicular polarization directions, do not induce the Dirac gap but modify the Dirac Fermi velocity. In the setup, graphene is on the $xy$ plane with $z=0$, and $x$ and $y$ are along the zigzag and armchair directions of the graphene structure, respectively. }
    \label{fig:illustration} 
\end{figure}

With the photon-free QED-HF formulation, we investigate graphene coupled with various types of cavity electromagnetic fields, and find a substantial renormalization of Dirac electronic states. 
Our results show that the photon-mediated interactions induce a topologically nontrivial Dirac gap in graphene for a circularly polarized cavity photon mode and a trivial Dirac gap with a flat band feature for a linearly polarized mode, as illustrated in Fig. \ref{fig:illustration}, with the latter arising purely from the photon-induced nonlocal interactions. 
The gap opening from a circularly polarized mode, originating from time-reversal symmetry breaking, has a classical analogue \cite{Floquet_graphene_Oka,Floquet_graphene_Fu}, while the gap opening from a linearly polarized mode, is a result of the anisotropic long-range interactions, and is a pure quantum effect induced by the cavity. This demonstrate a key difference in the phenomenology of cavity materials engineering vs Floquet engineering \cite{Hannes_NM_comment,Hannes_review_MQT}.
When two cavity photon modes are present, all symmetries of intrinsic graphene can be restored, resulting in a gapless Dirac state with a modified Fermi velocity. Moreover, by tuning the mode polarization, amplitude, and frequency, the Dirac cones in graphene can be flexibly renormalized, establishing material symmetry engineering via optical cavities as a powerful tool for the manipulation of electronic band structure and its topology. 
The developed QED-HF theoretical framework is general, and can be implemented as a first-principles method. As such, it is applicable to a wide range of materials and cavity configurations, and therefore opens the way to accurately model the electronic structure of materials strongly coupled to quantum vacuum field fluctuations without biasing the results through perturbative expansions.

\section{Non-Perturbative Theory for interacting electrons and photons}
Here we present the theoretical framework for the non-perturbative modeling of an electronic system interacting with the fluctuating photons in a cavity. As discussed below, the framework is built using a HF approach and relies on a photon-free QED treatment of the light-matter coupled problem.

\subsection{Photon-free QED Hamiltonian}
\label{sec:photon_free_QED}
For a system of $N_\text{e}$ (non-relativistic) electrons interacting with the electromagnetic field confined by a cavity, the minimal-coupling prescription in Coulomb gauge together with an effective cavity photon mode description in the long-wavelength approximation \cite{Mark_theory} results in the following Pauli-Fierz QED Hamiltonian \cite{Faisal_book}
\begin{equation}
\label{eq:QED_Hamiltonian}
\begin{aligned}
\hat{H} =\sum_i^{N_\text{e}}\left[\frac{(\hat{\mathbf{p}}_i-q \hat{\mathbf{A}})^2}{2 m}+\hat{V}\left(\mathbf{r}_i\right)\right]+\hbar \omega\left(\frac{1}{2}+\hat{a}^{\dagger} \hat{a}\right), \\
\end{aligned}
\end{equation}
where $\hat{\mathbf{p}}_i=-i\hbar\nabla_{\mathbf{r}_i}$ is the momentum operator for the electrons (with electron charge $q=-|e|$ and mass $m$),  and $\hat{\mathbf{A}}=A_0\left(\hat{a}^{\dagger}\mathbf{e}^*+\hat{a}\mathbf{e}\right)$ is the effective electromagnetic vector-potential operator with mode amplitude $A_0$ and polarization vector $\mathbf{e}$, and $\omega$ is the cavity mode frequency. Physically, the Hamiltonian consists of three parts: the bare electron part $\hat{H}_\text{e} = \sum_i^{N_\text{e}} \hat{h}_\text{e} (\mathbf{r}_i) \text{ with } \hat{h}_\text{e} (\mathbf{r}_i) =  \frac{\hat{\mathbf{p}}_i^2}{2 m}+\hat{V}\left(\mathbf{r}_i\right)$,  the bare photon part $\hat{H}_\text{p} = \hbar \omega\left(\frac{1}{2}+\hat{a}^{\dagger} \hat{a}\right)$, and the electron-photon interaction part $\hat{H}_\text{int}  =  -\frac{q}{m} \sum_i^{N_\text{e}} \hat{\mathbf{p}}_i \cdot \hat{\mathbf{A}} + \frac{q^2 N_\text{e} \hat{\mathbf{A}}^2}{2m}$. The interaction part is in turn composed of a paramagnetic term with $\sum_i^{N_\text{e}}\hat{\mathbf{p}}_i \cdot \hat{\mathbf{A}}$ and a diamagnetic term with $\hat{\mathbf{A}}^2$. 

The diamagnetic term can be absorbed into the bare effective photon term (see Appendix \ref{sec:appendix_modetrans}), reducing the Hamiltonian in Eq. \eqref{eq:QED_Hamiltonian} to
\begin{equation}
\label{eq:dressed_QED_Hamiltonian}
    \hat{H} = \hat{H}_\text{e} + \hbar \tilde{\omega} \left( \frac{1}{2} +\hat{\tilde{a}}^{\dagger} \hat{\tilde{a}} \right) - \frac{q}{m} \sum_i^{N_\text{e}} \hat{\mathbf{p}}_i \cdot \hat{\tilde{\mathbf{A}}} 
\end{equation}
with the dressed mode frequency $\tilde{\omega}$ and vector-potential operator $\hat{\tilde{\mathbf{A}}} = \tilde{A_0} \left(\hat{\tilde{a}}^{\dagger}\mathbf{e}^*+\hat{\tilde{a}} \mathbf{e}\right)$. The polarization of the dressed mode remains the same as that of the undressed mode. For a linearly polarized photon mode, diamagnetism renormalizes the mode frequency and amplitude to $\tilde{\omega} = \omega \sqrt{1+\zeta\frac{2N_\text{e} A_0^2}{\omega}}$ and $\tilde{A_0} = A_0 \frac{\sqrt{u+1} - \sqrt{u-1}}{\sqrt{2}}$ with  $\zeta = \frac{q^2}{m\hbar} $ $(=1 \text{ in atomic unit})$ and $u = \frac{\zeta N_\text{e} A_0^2 +  \omega}{\sqrt{ \left( 2 \zeta N_\text{e} A_0^2 +  \omega \right)  \omega }}$, respectively. Differently, for a circularly polarized photon mode, diamagnetism renormalizes the mode frequency and amplitude to $\tilde{\omega} = \omega \left( 1 + \zeta \frac{N_\text{e} A_0^2}{\omega} \right)$ and $\tilde{A_0} = A_0$, respectively.

The dressed Hamiltonian in Eq. \eqref{eq:dressed_QED_Hamiltonian} can be now represented in the dressed photon space $\{|\tilde{0}\rangle, \cdots, |\tilde{s} \rangle, \cdots\}$. Diagonal and subdiagonal matrix blocks have nonzero elements, and they satisfy $\langle \tilde{s} |\hat{H}| \tilde{s} \rangle = \langle \tilde{0}|\hat{H}| \tilde{0} \rangle + \tilde{s} \hbar \tilde{\omega}$ and $\langle \tilde{s}|\hat{H}|\tilde{s}+1\rangle = \sqrt{\tilde{s}+1} \langle \tilde{0} |\hat{H}|\tilde{1}\rangle$. This structure highlights that when the dressed photon frequency is much higher than the typical electronic transition frequencies taken into account, the states in different photon sectors go off resonance, so the Hamiltonian matrix can be conveniently downfolded into the zero photon sector and hence remove the photonic degrees of freedom \cite{Simone_PNAS,PhysRevLett.126.153603}. At the first order with respect to $\frac{\tilde{A}_0^2}{\tilde{\omega}}$, the downfolding gives rise to the effective photon-free QED Hamiltonian (see Appendix \ref{sec:appendix_downfolding})
\begin{equation}
\label{eq:effective_hamiltonian}
\begin{aligned}
    \hat{H}_\text{eff}  = \hat{H}_\text{e} + \frac{\hbar \tilde{\omega} }{2} + \hat{H}_\text{l} + \hat{H}_\text{nl}
\end{aligned},    
\end{equation}
where 
\begin{equation}
\label{eq:effective_hamiltonian_localpart}
    \hat{H}_\text{l} = - \zeta \frac{ \tilde{A}_0^{2}}{\tilde{\omega}} \sum_i^{N_\text{e}} \left( \hat{\mathbf{p}}_i \cdot \mathbf{e}\right) \left( \hat{\mathbf{p}}_i \cdot \mathbf{e}^* \right)
\end{equation}
represents the photon-induced local single-electron interaction since it involves the momentum of the same particle $i$, and 
\begin{equation}
\label{eq:effective_hamiltonian_nonlocalpart}
    \hat{H}_\text{nl} = - \zeta \frac{ \tilde{A}_0^{2}}{\tilde{\omega} } \sum_i^{N_\text{e}} \sum_{j \neq i}^{N_\text{e}} \left( \hat{\mathbf{p}}_i \cdot \mathbf{e}\right) \left( \hat{\mathbf{p}}_j \cdot \mathbf{e}^* \right)
\end{equation}
represents the photon-induced nonlocal electron-electron interaction involving two particles $i$ and $j$. The induced electron-electron  interaction is proportional to the electron momentum, 
which is different from the usual Coulomb electron-electron interaction that is proportional to $\frac{1}{|\mathbf{r}_i - \mathbf{r}_j|}$. It is important to note that the photon-induced nonlocal interaction depends on the amplitude, frequency, and polarization of a photon mode, highlighting its tunable nature for controlling matter properties.  

This cavity-mediated electron-electron interaction is a purely quantum phenomenon arising from the transverse component of the fluctuating photon fields, an effect that has no counterpart with classical light. 
To illustrate this explicitly, we use a coherent state of light, $
| \tilde{\lambda} (t)\rangle = e^{-\frac{|\tilde{\lambda}|^2}{2}} \sum_{\tilde{s}=0}^{+ \infty} \frac{\tilde{\lambda}^{\tilde{s}}}{\sqrt{\tilde{s}!}} e^{-i \tilde{\omega} (\tilde{s}+\frac{1}{2})t} | \tilde{s} \rangle $ satisfying $\hat{\tilde{a}} | \tilde{\lambda} (t)\rangle  = \tilde{\lambda} e^{-i \tilde{\omega} t} | \tilde{\lambda} (t)\rangle$, as a proxy of classical light. We project the electron-photon Hamiltonian in Eq. \eqref{eq:dressed_QED_Hamiltonian} onto this state, yielding
\begin{equation}
\label{eq:projection_coherent}
    \langle \tilde{\lambda} (t) | \hat{H} | \tilde{\lambda} (t)\rangle =  \hat{H}_\text{e} + \hbar \tilde{\omega} \left( \frac{1}{2} + | \tilde{\lambda} |^2 \right) - \frac{q}{m} \sum_i^{N_\text{e}} \hat{\mathbf{p}}_i \cdot \boldsymbol{\mathsf{A}}(t) 
\end{equation}
with $ \boldsymbol{\mathsf{A}}(t) = \tilde{A}_0 \left( \tilde{\lambda}^* e^{i \tilde{\omega} t} \mathbf{e}^* + \tilde{\lambda} e^{-i \tilde{\omega} t} \mathbf{e}\right)$. Through this projection, the vector-potential operator $\hat{\tilde{\mathbf{A}}}$ acts as a classical vector potential $\boldsymbol{\mathsf{A}}(t)$, leaving only light-mediated single electron corrections. The projected Eq. \eqref{eq:projection_coherent} is the Hamiltonian for classical time-periodic light driving in Floquet engineering \cite{Hannes_NM_comment}.

\subsection{Photon-free QED-HF formulation}
\label{sec:QED_HF}
To find the ground state of the many-particle Hamiltonian in Eq. \eqref{eq:effective_hamiltonian}, a HF approximation is applied to treat the photon-induced electron interactions in Eq. \eqref{eq:effective_hamiltonian_localpart} and \eqref{eq:effective_hamiltonian_nonlocalpart}. The electronic wavefunction is represented by a single Slater determinant that is constructed from the occupied single-particle orbitals $\{\varphi_i\}$ with energies $\{\varepsilon_i\}$. The orbitals can be obtained by solving the photon-free QED-HF equation (derived from the minimization of the total energy, see Appendix \ref{sec:appendix_HF}) 
\begin{equation}
\label{eq:hf_equation}
\begin{aligned}
\hat{\mathcal{F}} | \varphi_i  \rangle = \varepsilon_i | \varphi_i  \rangle \\
\end{aligned}
\end{equation}
with the Fock operator in first quantization
\begin{equation}
\label{eq:fock_operator}
\mathcal{F}(\hat{\mathbf{r}}) = h(\hat{\mathbf{r}})-\sum_{j}^{N_\text{e}} \mathcal{J}_j(\hat{\mathbf{r}})-\mathcal{K}_j(\hat{\mathbf{r}}),
\end{equation}
where the local part, consisting of the uncoupled electronic Hamiltonian and the operator from local interaction, is
\begin{equation}
\label{eq:local_hf_part}
h(\hat{\mathbf{r}})=h_\text{e}(\hat{\mathbf{r}})- \zeta \frac{ \tilde{A}_0^{2}}{\tilde{\omega}} \hat{\Pi}_{\text{l}}
\end{equation}
with $\hat{\Pi}_{\text{l}} = \left(\hat{\mathbf{p}}_{\mathbf{r}} \cdot \mathbf{e} \right)\left( \hat{\mathbf{p}}_{\mathbf{r}} \cdot \mathbf{e}^* \right)$, the direct operator from nonlocal interaction is 
\begin{equation}
\label{eq:nonlocal_direct}
\begin{aligned}
    \mathcal{J}_j(\mathbf{r}) f(\mathbf{r})=    \langle \varphi_j | \zeta \frac{ \tilde{A}_0^{2}}{\tilde{\omega}} \hat{\Pi}_{\text{nl}} | \varphi_j \rangle f(\mathbf{r})
\end{aligned},
\end{equation}
and the exchange operator from nonlocal interaction is
\begin{equation}
\label{eq:nonlocal_exchange}
    \begin{aligned}
        \mathcal{K}_j(\mathbf{r}) f(\mathbf{r})=  \langle \varphi_j | \zeta \frac{ \tilde{A}_0^{2}}{\tilde{\omega}} \hat{\Pi}_{\text{nl}} | f \rangle \varphi_j(\mathbf{r})
    \end{aligned} 
\end{equation}
with $\hat{\Pi}_{\text{nl}} = \left( \hat{\mathbf{p}}_{\mathbf{r}} \cdot \mathbf{e} \right)\left(\hat{\mathbf{p}}_{\mathbf{r}^{\prime}} \cdot \mathbf{e}^* \right)  + \left( \hat{\mathbf{p}}_{\mathbf{r}} \cdot \mathbf{e}^* \right)\left(\hat{\mathbf{p}}_{\mathbf{r}^{\prime}} \cdot \mathbf{e} \right)$ for any single-particle wavefunction $f(\mathbf{r})$. The integral in Eq. \eqref{eq:nonlocal_direct} and \eqref{eq:nonlocal_exchange} is for coordinate $\mathbf{r}^\prime$.  Because the Fock operator depends on its eigenstates, i.e., the occupied orbitals in Eq. \eqref{eq:hf_equation}, self-consistent calculations are required. While the self-consistent HF approach has been used for finite molecules in chemistry \cite{ab_HF_molecule}, only the non-self-consistent perturbative HF approach has been implemented for extended crystalline systems \cite{graphene_Vasil}.  

With spatial periodicity, the electrons in crystals outside a cavity are described by the Bloch states given by $h_\text{e} (\mathbf{r}) \varphi_{n\mathbf{k}}^0(\mathbf{r}) = \varepsilon_{n\mathbf{k}}^0 \varphi_{n\mathbf{k}}^0(\mathbf{r})$, where $n$ and $\mathbf{k}$ define the bands and crystal momenta, respectively. To numerically solve Eq. \eqref{eq:hf_equation}, the Fock operator $\hat{\mathcal{F}} $ is expressed as a matrix, which we refer to as photon-free QED Fock matrix, in the basis set constructed from the electronic states $\{\varphi_{n\mathbf{k}}^0 \}$ without interaction to cavity photons. The matrix elements 
$ {F}_{n \mathbf{k}~n' \mathbf{k}'} = \langle \varphi_{n \mathbf{k}}^{0}  | \hat{\mathcal{F}} | \varphi_{n^\prime \mathbf{k}^\prime}^{0}  \rangle = \delta_{\mathbf{k}\mathbf{k}^\prime} \langle \varphi_{n \mathbf{k}}^{0}  | \hat{\mathcal{F}} | \varphi_{n^\prime \mathbf{k}}^{0}  \rangle $ indicate that the matrix is diagonal in $\mathbf{k}$ space (see Appendix \ref{sec:appendix_Fock}). Thus, the QED Fock matrix can be constructed independently for each crystal momentum $\mathbf{k}$, i.e. ${F}_{nn'\mathbf{k}} = {F}_{\text{l},nn'\mathbf{k}} + {F}_{\text{nl},nn'\mathbf{k}}$. The contribution from the local operator is
\begin{equation}
\label{eq:matrixlocal}
\begin{aligned}
  {F}_{\text{l},n n' \mathbf{k}}  = \varepsilon_{n  \mathbf{k}}^0 \delta_{nn^\prime} -  \langle \varphi_{n \mathbf{k}}^{0}  | \zeta \frac{ \tilde{A}_0^{2}}{\tilde{\omega}} \hat{\Pi}_\text{l} | \varphi_{n^\prime \mathbf{k}}^{0}  \rangle
\end{aligned},
\end{equation}
and the contribution from the nonlocal operator is
\begin{equation}
\label{eq:matrixnonlocal}
    {F}_{\text{nl},n n' \mathbf{k}} = {J}_{n n' \mathbf{k}} + {K}_{n n' \mathbf{k}}
\end{equation}
with the direct component
\begin{equation}
\begin{aligned}
\label{eq:matrixdirect}
{J}_{n n' \mathbf{k}} =  -  \sum_{m \mathbf{k}^\prime}^{\text{occ}} \langle \varphi_{n \mathbf{k}}^{0}  \varphi_{m \mathbf{k}^\prime}  | \zeta \frac{ \tilde{A}_0^{2}}{\tilde{\omega}} \hat{\Pi}_\text{nl} | \varphi_{n^\prime \mathbf{k}}^{0}  \varphi_{m \mathbf{k}^\prime}   \rangle
\end{aligned}   
\end{equation}
and the exchange component
\begin{equation}
\label{eq:matrixexchange}
\begin{aligned}
{K}_{n n' \mathbf{k}} =   \sum_{m}^{\text{occ}} \langle \varphi_{n \mathbf{k}}^{0}  \varphi_{m \mathbf{k}}  | \zeta \frac{ \tilde{A}_0^{2}}{\tilde{\omega}} \hat{\Pi}_\text{nl} | \varphi_{m \mathbf{k}}  \varphi_{n^\prime \mathbf{k}}^{0}  \rangle
\end{aligned},    
\end{equation}
where $m$ indexes the occupied HF Bloch bands. 
The direct component $J_{n n^\prime \mathbf{k}}$ accounts for the interaction of an electron with $\mathbf{k}$ to all the other electrons and itself, due to the sum of occupied orbitals for all crystal momenta $\mathbf{k}^\prime$ in Eq. \eqref{eq:matrixdirect}. In contrast, the 
exchange component $K_{n n^\prime \mathbf{k}}$ accounts for the interaction of an electron with $\mathbf{k}$ to itself and the other electrons with the same crystal momentum $\mathbf{k}$, shown in Eq. \eqref{eq:matrixexchange}. 
Importantly, the self-direct and self-exchange interactions in direct and exchange components, respectively, exactly cancel with each other (see Appendix \ref{sec:appendix_HF}); as a result, the inclusion of both direct and exchange components guarantees the QED-HF approach to be self-interaction free.

The direct component $J_{n n^\prime \mathbf{k}}$ is proportional to the total electron momentum $\mathbf{P} = \sum_{m\mathbf{k}^\prime}^{\text{occ}} \langle \varphi_{m{\mathbf{k}^\prime}}  | \hat{\mathbf{p}}_\mathbf{r} | \varphi_{m{\mathbf{k}^\prime}}  \rangle$ of the  system. When $\mathbf{P} = 0$, the $J_{n n^\prime \mathbf{k}}$ is zero, meaning that the Fock matrix is only contributed by the local and exchange components; thus, the Fock matrix for each crystal momentum $\mathbf{k}$ depends only on the states with the same crystal momentum $\mathbf{k}$, which allows the separate self-consistent iterations of wavefunctions for each $\mathbf{k}$.
The resulting Fock matrix is solved and iteratively updated until convergence is reached.
 
In the local, direct and exchange components [Eq. \eqref{eq:matrixlocal}, \eqref{eq:matrixdirect} and \eqref{eq:matrixexchange}] of the photon-free QED Fock matrix, the interaction prefactor $\frac{\tilde{A}_0^{2}}{\tilde{\omega}}$  scales with the number of electrons $N_{\text{e}}$, such as $\frac{\tilde{A}_0^{2}}{\tilde{\omega}} = \frac{A_0^2}{\omega} \left( {1 + \zeta \frac{2 N_e A_0^2}{\omega}} \right)^{-1}$ and $\frac{\tilde{A}_0^{2}}{\tilde{\omega}} = \frac{A_0^2}{\omega} \left( {1 + \zeta \frac{ N_e A_0^2}{\omega}} \right)^{-1}$ for a linearly and circularly polarized photon mode, respectively. This indicates that the modifications of electronic states by cavity fluctuating photons  depend on the size of material systems. 
As shown in Ref. \cite{Mark_theory}, there is an intrinsic limit to the range over which electrons in a material are coupled to the effective cavity photon mode. Indeed, since the effective photon mode volume is finite, only the electrons within this volume should be counted in the light-matter interaction. This means that in the bulk limit ~\cite{Mark_theory}, the light-matter coupling is maximized. 
Filling the mode volume with matter has an effect of the photon dressing from diamagnetism, which in practice implies that when fabricating and designing the cavity the target frequency and amplitude of photon modes to be realized are the dressed ones. 
Finally, compared with exact diagonalization, the photon-free QED-HF approximation is equivalent to an expansion of the electron-photon coupling problem on the ground and singly-excited configuration states (see Appendix \ref{sec:appendix_Brillouin}).

\subsection{The case of multiple cavity photon modes}
A single effective photon mode inherently breaks the symmetry of matter by constraining the vacuum field to a single polarization. To include multiple polarizations, we allow for 
more cavity photon modes in the description of the light-matter interacting Hamiltonian.
For a multi-mode cavity, the vector-potential operator for $N_\text{p} > 1$ photons is $\hat{\mathbf{A}}= 
\sum_\alpha^{N_\text{p}} \hat{\mathbf{A}}_\alpha = \sum_{\alpha}^{N_\text{p}}  A_{0 \alpha} \left( \hat{a}_{\alpha}^{\dagger} \mathbf{e}_{\alpha}^{*} + \hat{a}_{\alpha} \mathbf{e}_{\alpha} \right)$ with the mode index $\alpha$. While the diamagnetic term resulting from this vector-potential operator might mix different photons, a normal mode transformation $ \hat{\mathbf{A}}= 
\sum_\alpha^{N_\text{p}} \hat{\mathbf{A}}_\alpha \rightarrow \hat{\tilde{\mathbf{A}}}= 
\sum_\alpha^{N_\text{p}} \hat{\tilde{\mathbf{A}}}_\alpha$ can be found \cite{Faisal_book}, so that the multi-mode electron-photon Hamiltonian can be written as
\begin{equation}
\label{eq:qed_hamiltonian_multimodes}
    \hat{H} = \hat{H}_\text{e} + \sum_\alpha^{N_\text{p}} \hbar \tilde{\omega}_\alpha \left( \frac{1}{2} +\hat{\tilde{a}}_\alpha^{\dagger} \hat{\tilde{a}}_\alpha \right) - \frac{q}{m} \sum_\alpha^{N_\text{p}} \sum_i^{N_\text{e}} \hat{\mathbf{p}}_i \cdot \hat{\tilde{\mathbf{A}}}_{\alpha}
\end{equation}
with the dressed mode frequency $\tilde{\omega}_\alpha$, amplitude $\tilde{A}_{0\alpha}$ and polarization $\tilde{\mathbf{e}}_\alpha$. 

A high-frequency downfolding procedure, equivalent to the one for single mode, can be applied to the multi-mode Hamiltonian in Eq. \eqref{eq:qed_hamiltonian_multimodes}. The multi-mode effective photon-free Hamiltonian is then given by (see Appendix \ref{sec:appendix_downfolding}) 
\begin{equation}
\label{eq:multimode_effective_hamiltonian}
    \hat{H}_\text{eff}  = \hat{H}_\text{e} + \sum_\alpha^{N_\text{p}} \left( \frac{\hbar \tilde{\omega}_\alpha}{2} + \hat{H}_{\text{l},\alpha} + \hat{H}_{\text{nl},\alpha} \right)
\end{equation}
with $\hat{H}_{\text{l},\alpha} = - \zeta \frac{ \tilde{A}_{0\alpha}^{2}}{\tilde{\omega}_\alpha} \sum_i^{N_\text{e}} \left( \hat{\mathbf{p}}_i \cdot \tilde{\mathbf{e}}_\alpha \right) \left( \hat{\mathbf{p}}_i \cdot \tilde{\mathbf{e}}_\alpha^* \right)$ and $\hat{H}_{\text{nl},\alpha} = - \zeta \frac{ \tilde{A}_{0\alpha}^{2}}{\tilde{\omega}_\alpha} \sum_i^{N_\text{e}} \sum_{j \neq i}^{N_\text{e}} \left( \hat{\mathbf{p}}_i \cdot \tilde{\mathbf{e}}_\alpha \right) \left( \hat{\mathbf{p}}_j \cdot \tilde{\mathbf{e}}_\alpha^* \right)$.  This shows that the photon-free Hamiltonian in the case of multiple photons is a summation of the Hamiltonians corresponding to the demixed normal effective modes. 

The photon-free QED-HF formulation in Sec. \ref{sec:QED_HF} can be directly extended to the cavities with multiple effective photon modes ($N_\text{p} > 1$). Since the photon-free Hamiltonian in Eq. \eqref{eq:multimode_effective_hamiltonian} features a summation of electron interactions from all the normal effective photon modes, accordingly, the operator $\frac{\tilde{A}_0^{2}}{\tilde{\omega}} \hat{\Pi}_\text{l}$ in Eq. \eqref{eq:local_hf_part} for a single mode should be replaced by $\sum_\alpha^{N_\text{p}} \frac{\tilde{A}_{0\alpha}^{2}}{\tilde{\omega}_\alpha} \hat{\Pi}_{\text{l}, \alpha}$ with $\hat{\Pi}_{\text{l},\alpha} = \left(\hat{\mathbf{p}}_{\mathbf{r}} \cdot \tilde{\mathbf{e}}_\alpha \right)\left( \hat{\mathbf{p}}_{\mathbf{r}} \cdot \tilde{\mathbf{e}}_\alpha^* \right)$, and the operator $\frac{\tilde{A}_0^{2}}{\tilde{\omega}} \hat{\Pi}_\text{nl}$ in Eq. \eqref{eq:nonlocal_direct} and \eqref{eq:nonlocal_exchange} by $\sum_\alpha^{N_\text{p}} \frac{\tilde{A}_{0\alpha}^{2}}{\tilde{\omega}_\alpha} \hat{\Pi}_{\text{nl}, \alpha}$ with $\hat{\Pi}_{\text{nl},\alpha} = \left( \hat{\mathbf{p}}_{\mathbf{r}} \cdot \tilde{\mathbf{e}}_\alpha \right)\left(\hat{\mathbf{p}}_{\mathbf{r}^{\prime}} \cdot \tilde{\mathbf{e}}_\alpha^* \right)  + \left( \hat{\mathbf{p}}_{\mathbf{r}} \cdot \tilde{\mathbf{e}}_\alpha^* \right)\left(\hat{\mathbf{p}}_{\mathbf{r}^{\prime}} \cdot \tilde{\mathbf{e}}_\alpha \right)$. Similar operator replacements should be applied to Eq. \eqref{eq:matrixlocal}, \eqref{eq:matrixdirect} and \eqref{eq:matrixexchange}. Note that summing the interactions over multiple modes can effectively enhance the interaction strength.

\section{Effect of cavity-mediated nonlocal electron interactions on Dirac states in graphene}
In this section, the photon-free QED-HF approach is employed to study the modifications of Dirac states in graphene by cavity photons, with a specific focus on the modifications resulting from the cavity-mediated nonlocal electron-electron interaction. 

The gapless Dirac states in graphene arise from the $2p_z$ orbitals of carbon atoms located at the $A$ and $B$ sites of a honeycomb lattice, satisfying, among others, time-reversal, spatial-inversion, and three-fold rotational symmetries. Graphene is described by a tight-binding model with a nearest-neighbour (NN) hopping energy of $t_0 = -2.7$ eV \cite{graphene_tb}, giving rise to Dirac cones with a Fermi velocity of $v_\text{F}=0.87\times10^6$ m/s at $\pm \mathbf{K}$ valleys. The main ingredient to calculate the cavity-mediated electron interactions in graphene is the momentum matrix element $\mathbf{p}_{mn\mathbf{k}}= -i \hbar \langle \varphi_{m\mathbf{k}}^0 | \nabla_\mathbf{r} | \varphi_{n\mathbf{k}}^0  \rangle $ with band indexes $m,n=\{v,c\}$ for lower valence ($v$) and upper conduction ($c$) Dirac bands with crystal momentum $\mathbf{k}$, whose evaluation from the contribution of $2p_z$ orbital \cite{graphene_momentum_fullTB,tight_binding_p_k} is described in Appendix \ref{sec:appendix_grapheneTB}.

In the ground state, the charge neutral graphene interacting with cavity photon modes has total momentum $\mathbf{P} = 0$. For each crystal momentum $\mathbf{k}$, the $2 \times 2$ QED Fock matrix in the basis set $\{ \varphi_{v \mathbf{k}}^0 , \varphi_{c \mathbf{k}}^0  \}$ is iteratively constructed by following Eq. \eqref{eq:matrixlocal}-\eqref{eq:matrixexchange} (see Appendix \ref{sec:appendix_grapheneFock}), and solved to find the cavity-renormalized HF orbitals.
The photon-free QED-HF solutions are converged to an accuracy of $10^{-12}$ eV in the HF energy eigenvalue.

The following subsections present the cavity-renormalized Dirac states in graphene, interacting with linearly and circularly polarized cavity photon modes, as illustrated in Fig. \ref{fig:illustration}. The frequency, amplitude and polarization of dressed effective normal photon modes are denoted as $\omega$, $A_0$ and $\mathbf{e}$, respectively, without tilde symbol (\textasciitilde) in the notation for brevity in this section.

\subsection{Circularly polarized photon}
\label{sec:a_circularly_polarized_photon}

\begin{figure}[!b] 
    \centering
    \includegraphics[width=\linewidth]{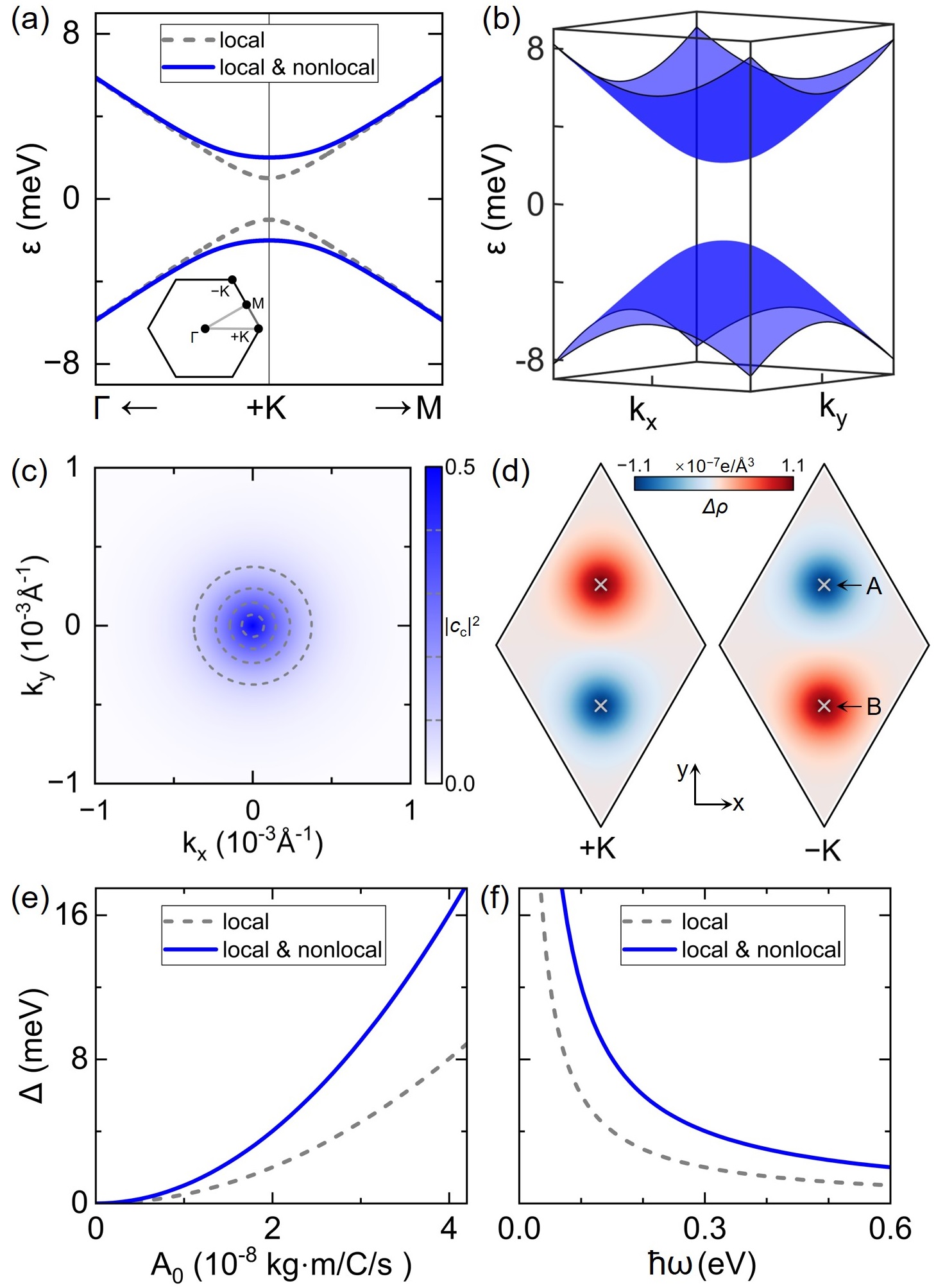}
    \caption{Dirac states in graphene coupled with a circularly polarized photon mode with $\hbar\omega=0.3$ eV, $A_0=2\times10^{-8}\frac{\text{kg} \cdot \text{m}}{\text{C} \cdot \text{s}}$, and $\mathbf{e}=\mathbf{e}_x+i\mathbf{e}_y$. 
    (a) HF bands of the $+\mathbf{K}$ valley (inset) along $\mathbf{\Gamma} \leftarrow +\mathbf{K} \rightarrow \mathbf{M}$ path. 
    (b) Representation of the blue band in (a) in the two-dimensional reciprocal zone $\{k_x , k_y\} \in [-1,1]$~$10^{-3}$\text{\AA}$^{-1}$ centered at the crystal momentum $+\mathbf{K}$. 
    (c) Component $|c_c|^2$ of the conduction basis state $\varphi_{c\mathbf{k}}^0$ for the lower valence band ($\varphi_{v\mathbf{k}} = c_v \varphi_{v\mathbf{k}}^0 + c_c \varphi_{c\mathbf{k}}^0$) in (b). 
    (d) Variation of the electron density, $\Delta \rho (\mathbf{r}) = \sum_\mathbf{k} |\varphi_{v \mathbf{k}}|^2 - |\varphi_{v \mathbf{k}}^0|^2$, at the specific $z = 0.33~\text{\AA}$ plane, where the $2p_z$ atomic orbital of carbon has its maximum. The left (right) panel shows the contribution to density from the $+\mathbf{K}$ ($-\mathbf{K}$) valley. The cross signs mark the position of the $A$ and $B$ sites.
    (e,f) Evolution of the Dirac band gap in panel (a) as a function of the $A_0$ (with fixed $\hbar\omega=0.3$ eV) and photon energy $\hbar\omega$ (with fixed $A_0 = 2 \times 10^{-8} \frac{\text{kg} \cdot \text{m}}{\text{C} \cdot \text{s}}$), respectively.}
    \label{fig:circular}    
\end{figure}

To describe the interaction between the electrons in graphene and a circularly polarized cavity photon mode, we set the polarization vector to $\mathbf{e} = \mathbf{e}_x + i \mathbf{e}_y $ (parallel to the graphene plane), the photon energy to $\hbar\omega=0.3$ eV, and the amplitude to $A_0 = 2 \times 10^{-8} \frac{\text{kg} \cdot \text{m}}{\text{C} \cdot \text{s}}$, which 
satisfies the high-frequency condition for the photon-free QED formulation for electrons close to the Dirac points. We find that the photon-induced local interaction opens an energy gap at the Dirac points as previously reported in literature \cite{graphene_Russian,graphene_Sentef,graphene_Vasil,graphene_cavity_topology_single_particle}, while the nonlocal interaction further increases the size of the gap, a quantum effect which has thus far not been discussed. As shown in Fig. \ref{fig:circular}(a), the local interaction breaks the degeneracy of Dirac point at $+\mathbf{K}$ valley, resulting in the massive Dirac cone with a band gap $\Delta \sim 2$ meV. With the inclusion of photon-induced nonlocal interaction, the Dirac gap is enlarged to $\Delta \sim 4$ meV [Fig. \ref{fig:circular}(a,b)]. This demonstrates that the photon-induced nonlocal electron-electron interaction in Eq. \eqref{eq:effective_hamiltonian_nonlocalpart}, originating from the quantum nature of cavity photons, plays an important role in the renormalization of Dirac states in graphene and can only be captured by a non-perturbative theoretical approach. 

To analyze the formation of massive Dirac cones induced by a circularly polarized cavity photon, we evaluate the electronic wavefunction and the change of electron density in reciprocal and real space, respectively. In the basis set $\{ \varphi_{v \mathbf{k}}^0 , \varphi_{c \mathbf{k}}^0  \}$, the wavefunction of the cavity-renormalized valence Dirac band in Fig. \ref{fig:circular}(b) is expressed as $\varphi_{v \mathbf{k}}  = c_v \varphi^{0}_{v \mathbf{k}}  + c_c \varphi^{0}_{c \mathbf{k}} $ for a given crystal momentum $\mathbf{k}$. Figure \ref{fig:circular}(c) shows $|c_c|^2$, i.e., the component of the original conduction Dirac state $\varphi_{c\mathbf{k}}^0 $ without interaction to photon. This indicates that the original valence $\varphi_{v \mathbf{k}}^0 $ and conduction $\varphi_{c \mathbf{k}}^0 $ states are hybridized by the circularly polarized cavity photon mode, leading to the formation of a massive Dirac cone at $+\mathbf{K}$ valley [Fig. \ref{fig:circular}(a,b)]. The hybridization is isotropic in reciprocal space, as expected from the symmetry of the electron-photon coupling, and it is worth mentioning that the equivalent behaviour is observed for the $-\mathbf{K}$ valley.

Figure \ref{fig:circular}(d) shows the cavity-induced modification of the electron density, $\Delta \rho (\mathbf{r}) = \sum_\mathbf{k} |\varphi_{v \mathbf{k}}|^2 - |\varphi_{v \mathbf{k}}^0|^2$, contributed from the reciprocal zone $\{k_x , k_y\} \in [-1,1]~10^{-3} \text{\AA}^{-1}$ with respect to crystal momenta $\pm \mathbf{K}$. The zone is large enough to contain all the modified Bloch states at $\pm \mathbf{K}$ valleys for obtaining the density variation (see Appendix \ref{sec:appendix_graphenedensity} for the calculation of electron density). For the $+ \mathbf{K}$ valley, the density increases at $A$ sites, and decreases at $B$ sites. Oppositely, for the $-\mathbf{K}$ valley, the density decreases at $A$ sites, and increases at $B$ sites. As a result, the electron density from $\pm \mathbf{K}$ valleys is different $|\varphi_{v, +\mathbf{K}} (\mathbf{r})|^2 \neq |\varphi_{v, -\mathbf{K}} (\mathbf{r})|^2$, indicating that time-reversal symmetry is broken by the circularly polarized photon.

The size of the cavity-induced Dirac gap can be tuned by adjusting the amplitude and frequency of the circularly polarized photon. As shown in Fig. \ref{fig:circular}(e), for a fixed photon energy, the Dirac gap becomes larger with increasing mode amplitude, following the relation $\Delta \propto A_0^2$. For a fixed mode amplitude, instead, the Dirac gap becomes smaller with increasing mode frequency and evolves as $\Delta \propto \omega^{-1}$ [Fig. \ref{fig:circular}(f)]. Importantly, the 
overall evolution of the gap size follows $\Delta = \kappa \frac{A_0^2}{\omega}$. Although the evolution is similar to what is predicted for the Floquet Dirac gap in graphene induced by a time-periodic circularly polarized light field \cite{Floquet_graphene_Oka,Floquet_graphene_Fu}, the Floquet gap is only contributed from the local electron interaction mediated by the classical light. The dependence on the mode amplitude and frequency originates from the interaction prefactor $\frac{A_0^{2}}{\omega}$ in Eq. \eqref{eq:effective_hamiltonian_localpart} and \eqref{eq:effective_hamiltonian_nonlocalpart}, and the coefficient $\kappa$ is determined by the polarization dependent momentum matrix elements of graphene. As a sanity check we note that the gap size is on the order of $10$ meV, while the cavity photon energy on the order of $0.3$ eV, hence the high-frequency approximation for the photon-free QED Hamiltonian in Eq. \eqref{eq:effective_hamiltonian} is justified. 

A gap size of $\Delta = 1.8$ meV, induced by a circularly polarized cavity photon with $\hbar\omega =0.3$ eV and 
$A_0 = 1.87 \times 10^{-8} \frac{\text{kg} \cdot \text{m}}{\text{C} \cdot \text{s}}$ has been previously reported based on perturbation theory \cite{graphene_Vasil, graphene_Sentef,graphene_Russian}. This is quantitatively consistent with our results for the case with only the photon-mediated local electron interaction, which demonstrates that the photon-induced nonlocal electron-electron interaction is missed in the previous perturbation modeling. Thus, the non-perturbative self-consistent photon-free QED-HF approach is necessary for the predictions of the photon-induced nonlocal interaction on the renormalization of Dirac states in graphene, as a unique quantum effect arising from cavity photons (with no counterpart for classical light, as discussed in Sec. \ref{sec:photon_free_QED}).  

\subsection{Linearly polarized photon}
\label{sec:a_linearly_polarized_photon}
\begin{figure}[!ht]
    \centering
    \includegraphics[width=1\linewidth]{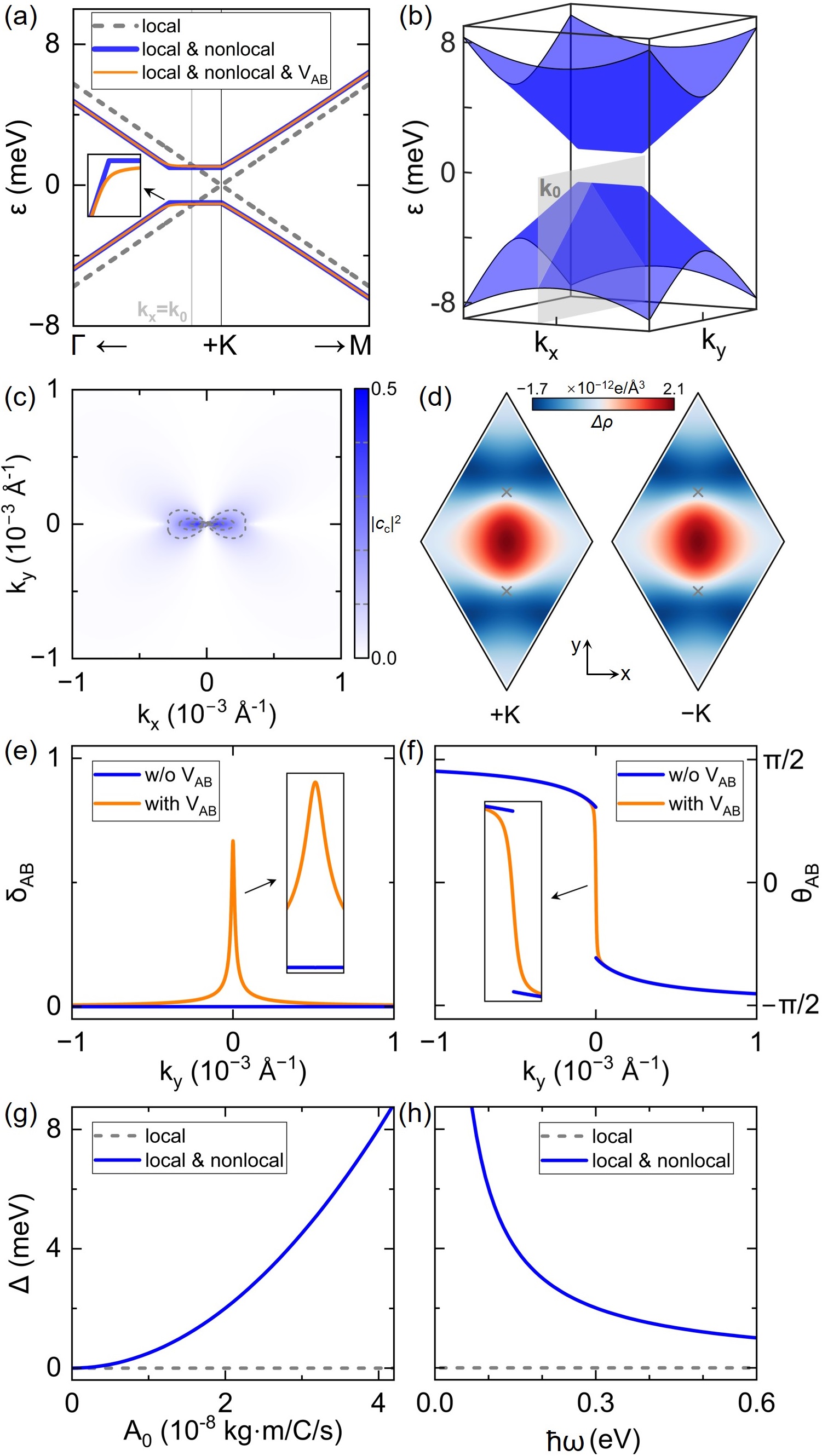}
    \caption{Dirac states in graphene coupled with a linearly polarized photon mode with $\hbar\omega=0.3~\text{eV}$, $A_0 = 2 \times 10^{-8} \frac{\text{kg} \cdot \text{m}}{\text{C} \cdot \text{s}}$, and $\mathbf{e} = \mathbf{e}_x$. 
    (a) HF bands of the $+\mathbf{K}$ valley with and without nonlocal interaction. The former are also shown in the presence of a tiny sublattice potential value $V_{AB} = \pm 2 \times 10^{-5} t_0$. 
    (b) Representation of the blue band in (a) in $\{k_x , k_y\} \in [-1,1]$~$10^{-3}$\text{\AA}$^{-1}$ centered at $+\mathbf{K}$. 
    (c) Component $|c_c|^2$ of the conduction basis state $\varphi_{c\mathbf{k}}^0$ for the lower valence band ($\varphi_{v\mathbf{k}} = c_v \varphi_{v\mathbf{k}}^0 + c_c \varphi_{c\mathbf{k}}^0$) in (b).
    (d) Variation of electron density, $\Delta \rho (\mathbf{r}) = \sum_\mathbf{k} |\varphi_{v \mathbf{k}}|^2 - |\varphi_{v \mathbf{k}}^0|^2$, at $z = 0.33~\text{\AA}$ plane for $V_{AB} = 0$. The left (right) panel shows the contribution from the $+\mathbf{K}$ ($-\mathbf{K}$) valley.  
    (e,f) The difference of wavefunction amplitude $\delta_{AB}$ (phase $\theta_{AB}$) between the $A$ and $B$ sites for the valence states across the plane $k_x = k_0 = -2 \times 10^{-4}~\text{\AA}^{-1}$ (the gray plane in (b)). Blue (Orange) lines are for graphene with $V_{AB} = 0$ ($\pm 2 \times 10^{-5} t_0$). 
    (g,h) Evolution of the Dirac band gap in panel (a) as a function of $A_0$ (with fixed $\hbar\omega=0.3$ eV) and $\hbar\omega$ (with fixed $A_0 = 2 \times 10^{-8} \frac{\text{kg} \cdot \text{m}}{\text{C} \cdot \text{s}}$) for $V_{AB} = 0$, respectively.}
    \label{fig:linear}
\end{figure}

For a linearly polarized photon mode, we choose the polarization $\mathbf{e} = \mathbf{e}_x$ (i.e. along the zigzag direction of graphene), and keep the photon energy $\hbar\omega = 0.3 $ eV and mode amplitude $A_0 = 2 \times 10^{-8} \frac{\text{kg} \cdot \text{m}}{\text{C} \cdot \text{s}}$ as before. As shown in Fig. \ref{fig:linear}(a), the photon-mediated local interaction does not break the Dirac degeneracy. In contrast, with the induced nonlocal interaction, the Dirac degeneracy is destroyed, and a band gap of $\Delta \sim 2$ meV is formed at both $\pm \mathbf{K}$ valleys. The gapped Dirac cones are anisotropic in reciprocal space, as shown in Fig. \ref{fig:linear}(b).

Different from the case of a circularly polarized photon in Sec. \ref{sec:a_circularly_polarized_photon} where the renormalization is isotropic, the cavity-renormalized Dirac states by a linearly polarized photon are in an anisotropic superposition of valence and conduction basis states in reciprocal space. This can be quantified through the electron wavefunction $\varphi_{v \mathbf{k}}  = c_v \varphi^{0}_{v \mathbf{k}}  + c_c \varphi^{0}_{c \mathbf{k}} $ shown in Fig. \ref{fig:linear}(c). Here, the conduction component $|c_c|^2$ has a butterfly shape elongated along the $k_x$ direction in reciprocal space, indicating that three-fold rotational symmetry is broken by the linearly polarized photon.

Figure \ref{fig:linear}(d) shows the modification of the electron density from the $\pm \mathbf{K}$ valleys, where the density variation, which is the same for the two valleys, is mainly on the $A\text{-}B$ bonds between the NN sites. The density increases on the vertical bonds, and decreases on the other bonds, showing that the three-fold rotational symmetry, intrinsic to the bare graphene, is broken. At the same time, the renormalized Dirac electron wavefunctions satisfy $|\varphi_{v, +\mathbf{K}} (\mathbf{r})|^2 = |\varphi_{v, -\mathbf{K}} (\mathbf{r})|^2$, indicating that time-reversal symmetry is not broken. In other words, the Dirac gap opening in the case of a linearly polarized cavity photon is not a consequence of time-reversal symmetry breaking, and instead connected to the long-range anisotropy (characterized by three-fold rotational symmetry breaking) in the presence of the photon-induced nonlocal electron-electron interaction.

In Fig. \ref{fig:linear}(a,b), approaching $+\mathbf{K}$ along the $+\mathbf{K} \text{-} \mathbf{\Gamma}$ direction, the renormalized Dirac bands change abruptly to become perfectly flat. Perpendicular to the flat line, the bands are instead shaped like a wedge. As discussed below in Sec. \ref{sec:effective_hopping_integrals}, these sharp kink features are a consequence of an unphysical infinitely long-range electron-electron interaction due to the long wavelength approximation, and can be prevented by truncating the interaction range. Interestingly, by introducing a small sublattice potential difference $V_{AB} = \pm 2 \times 10^{-5} t_0$ at $A$ and $B$ sites (or a weak spin-orbit coupling strength, which would be there for realistic graphene), the bands smoothen, as shown in Fig. \ref{fig:linear}(a). The long-range anisotropy from nonlocal interaction dominates the global band renormalization as in the case of $V_{AB} = 0$ (see Appendix \ref{sec:appendix_grapheneFock} and Sec. \ref{sec:effective_hopping_integrals}).

To understand the flat-line band, we analyze the wavefunctions $\varphi_{v\mathbf{k}} = c_{A} \varphi_{A\mathbf{k}}^{0} + c_{B} \varphi_{B\mathbf{k}}^{0}$ for valence Dirac states in the sublattice basis set $\{\varphi_{A\mathbf{k}}^{0}, \varphi_{B\mathbf{k}}^{0}\}$ (see Appendix \ref{sec:appendix_grapheneTB} for the $\varphi_{A\mathbf{k}}^{0}$ and $\varphi_{B\mathbf{k}}^{0}$ functions). We specifically choose a path along $k_y$ for $k_x = k_0 = -2 \times 10^{-4}$\AA$^{-1}$ [indicated by the gray plane in Fig. \ref{fig:linear}(b)]. We consider both the case of zero and finite sublattice potential $V_{AB}$. In Fig. \ref{fig:linear}(e), we plot the quantity $\delta_{AB} = |c_{A\mathbf{k}}| - |c_{B\mathbf{k}}|$, which describes the difference in the magnitudes of the wavefuction coefficients for the two lattice sites. When $V_{AB}=0$, both sites are equivalent, i.e., $\delta_{AB} = 0$, preserving spatial inversion symmetry. Looking at the phase difference $\theta_{AB} = \text{Arg}(c_{A\mathbf{k}}) - \text{Arg}(c_{B\mathbf{k}})$ in Fig. \ref{fig:linear}(f), we observe a discontinuity crossing the flat line, where a QED-HF solution can not be found exactly at the kink and flat line. This indicates the presence of an unphysical singularity. The singularity is removed by introducing a finite $V_{AB}$, as evidenced by the continuous and smooth behaviors of $\delta_{AB}$ and $\theta_{AB}$ in Fig. \ref{fig:linear}(e) and \ref{fig:linear}(f). 

Finally, the size of the Dirac gap, induced by a linearly polarized photon, is also tunable through adjustments of the photon parameters. As shown in Fig. \ref{fig:linear}(g,h), with only the local interaction, the Dirac cones remain gapless regardless of the choice of photon parameters. With the nonlocal electron-electron interaction included, the Dirac gap goes again as $\Delta = \xi \frac{A_0^2}{\omega}$, but the prefactor $\xi$ is different from that for a circularly polarized mode. Also, the length of the flat line is affected by photon parameters, and evolves as $L = \chi \frac{A_0^2}{\omega}$. The direction of the singular flat line in reciprocal space corresponds to the polarization direction of the photon mode in real space, and hence can be rotated in the $k_x k_y$ plane at will by changing the mode polarization in the $xy$ plane (see Appendix \ref{sec:appendix_rotation}). In Appendix \ref{sec:appendix_analyticalHF}, the emergence of the singular flat-line band feature is analytically demonstrated using the two-dimensional low-energy effective Dirac model, also the prefactors $\xi$ and $\chi$ for the bands evolution are analytically derived. The case of a linearly polarized cavity photon mode demonstrates that the non-perturbative QED-HF approach is essential to grasp the unique quantum effect, missed in perturbation theory, from the photon-induced nonlocal interaction in the collective electron-photon hybrid system. 

\subsection{Ellipticity and band topology}
\label{sec:evolution_and_band_topology}
\begin{figure}[b]
    \centering
    \includegraphics[width=1.02\linewidth]{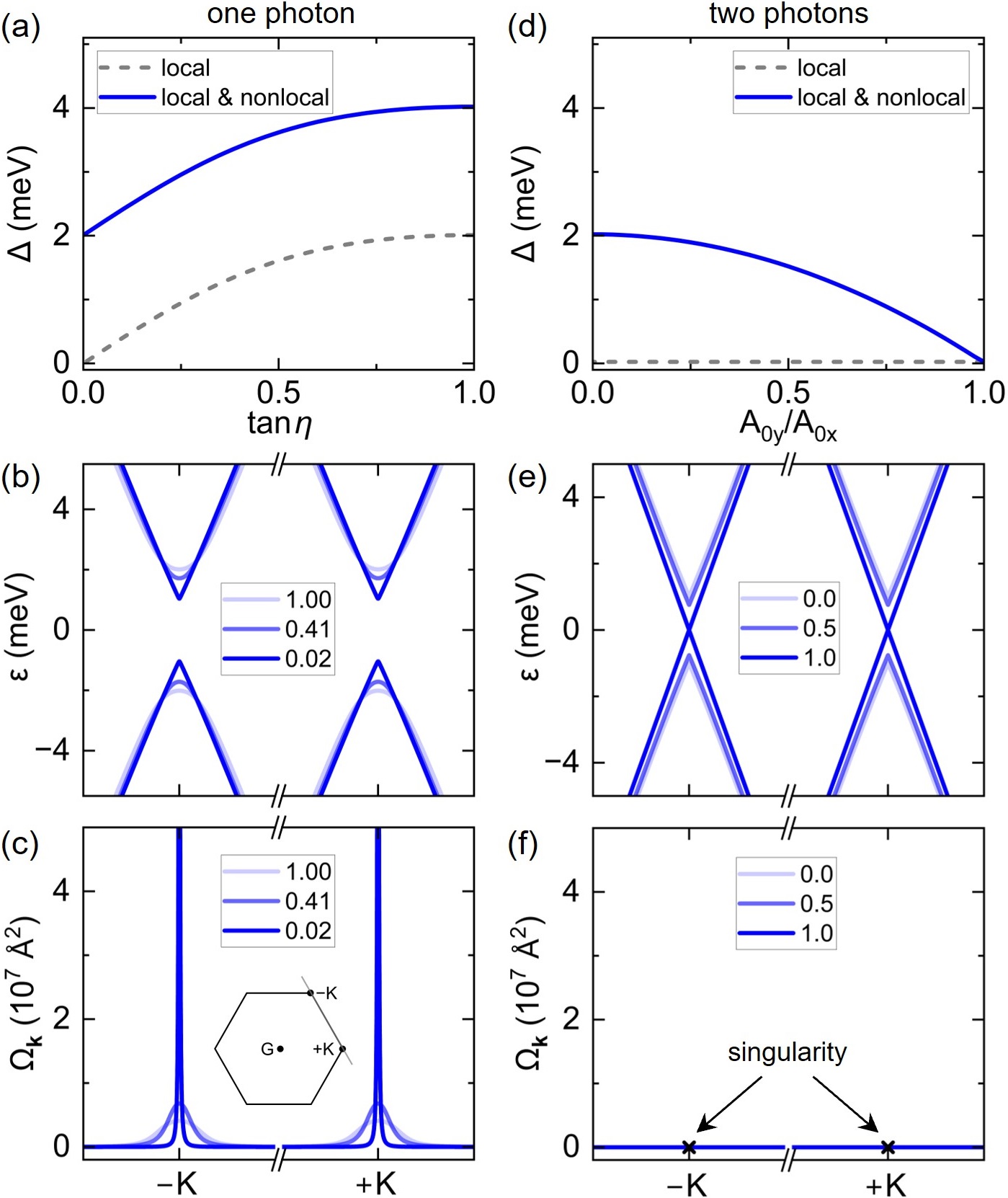}
    \caption{Evolution of cavity-renormalized Dirac states and band topology.  
    (a) Using a single photon mode with $\hbar\omega=0.3$ eV, $A_0 = 2 \times 10^{-8} \frac{\text{kg} \cdot \text{m}}{\text{C} \cdot \text{s}}$, and $\mathbf{e} = \mathbf{e}_x \cos{\eta}  + i \mathbf{e}_y \sin{\eta} $, the Dirac band gap changes with mode ellipticity degree $\tan{\eta}$.  
    (b) HF bands of the $\pm \mathbf{K}$ valleys, from local and nonlocal interactions, for ellipticity degrees $\tan{\eta} = 1, 0.41~\text{and}~0.02$, corresponding to ellipticity angles $\eta = \frac{\pi}{4}$, $\frac{\pi}{8}$ and $\frac{\pi}{180} $, respectively. 
    (c) Berry curvature for the valence band in (b). 
    (d) Change of the Dirac band gap with mode amplitude ratio $\frac{A_{0y}}{A_{0x}}$ in the presence of two linearly polarized photon modes with $\mathbf{e} = \mathbf{e}_x$ and $\mathbf{e}_y$. The photon energy of the two modes is $\hbar\omega=0.3$ eV, and the amplitude of $x$-polarized mode is fixed as $A_{0x} = 2 \times10^{-8} \frac{\text{kg} \cdot \text{m}}{\text{C} \cdot \text{s}}$.
    (e) HF bands of $\pm \mathbf{K}$ velleys, from local and nonlocal interactions, for the amplitude ratio $\frac{A_{0y}}{A_{0x}} = 0.0, 0.5,~\text{and}~1.0~$. 
    (f) Berry curvature for the valence band in (e), where the cross signs indicate that the Berry curvature is singular for the valence band at the crystal momenta $\pm \mathbf{K}$ .}
    \label{fig:topology}
\end{figure}

Here we investigate the renormalization of Dirac states by a cavity photon as its polarization transitions from circular to linear.  The elliptically polarized photon is implemented by setting the polarization vector to $\mathbf{e} = \mathbf{e}_x\cos{\eta}  + i \mathbf{e}_y\sin{\eta} $, and keeping the mode frequency and amplitude as before. A circularly and linearly polarized photon mode has the ellipticity degree $\tan{\eta} = 1$ and $0$ with the ellipticity angle $\eta = \frac{\pi}{4}$ and $0$, respectively. 
As shown in Fig. \ref{fig:topology}(a), for a single photon mode, the cavity-induced Dirac gap decreases with decreasing $\tan{\eta}$: with only the photon-mediated local interaction, the Dirac gap decreases to zero, whereas the gap stays finite with the nonlocal interaction included. We also note that decreasing the ellipticity degree, i.e. $\tan{\eta} = 1, 0.41, 0.02$ respectively with $\eta = \frac{\pi}{4}, \frac{\pi}{8}, \frac{\pi}{180}$,  the Dirac dispersion becomes sharper around crystal momenta $\pm \mathbf{K}$ [Fig. \ref{fig:topology}(b)]. 

Since a circularly polarized light field can affect the band topology of Dirac states in graphene \cite{graphene_Sentef, graphene_Vasil, graphene_cavity_topology_single_particle, Floquet_graphene_Oka,Floquet_graphene_Fu}, we now discuss the evolution of the Berry curvature of cavity-renormalized Dirac bands as the ellipticity degree $\tan{\eta}$ goes from 1 to 0. The Berry curvature for the occupied valence Dirac band is computed, after the self-consistent QED-HF iterations, as
\begin{equation}
\Omega_{\mathbf{k}}=-\frac{2 \operatorname{Im}\left[\left\langle\varphi_{v\mathbf{k}}\left|\hat{v}_x\right| \varphi_{c\mathbf{k}} \right\rangle\left\langle\varphi_{c\mathbf{k}}\left|\hat{v}_y\right| \varphi_{v\mathbf{k}}\right\rangle\right]}{\left(\varepsilon_{c\mathbf{k}}-\varepsilon_{v\mathbf{k}}\right)^2}
\end{equation}
with velocity operators $\hat{v}_x=\frac{\partial \hat{\mathcal{F}}}{\partial k_x}$ and $\hat{v}_y=\frac{\partial \hat{\mathcal{F}}}{\partial k_y}$ and the photon-free QED Fock operator $\hat{\mathcal{F}}$. Figure \ref{fig:topology}(c) shows that, with circular polarization ($\tan{\eta} = 1$), the $\Omega_{\mathbf{k}}$ for the valence band is non-zero around the $\pm \mathbf{K}$ valleys. The integration of $\Omega_\mathbf{k}$ in the first Brillouin zone (BZ) gives rise to a finite Chern invariant $C=\frac{1}{2 \pi} \int_\text{BZ} \Omega_{\mathbf{k}} d \mathbf{k}  =1$. This indicates a circularly polarized photon induces a topologically nontrivial phase, which can support quantum anomalous Hall states, and is consistent with time-reversal symmetry breaking shown in Sec. \ref{sec:a_circularly_polarized_photon}. 

As the ellipticity degree $\tan{\eta}$ decreases, the Berry curvature gets sharper around the crystal momenta $\pm \mathbf{K}$, as shown in Fig. \ref{fig:topology}(c). The Chern number is still $C = 1$, and hence the nontrivial topology is preserved for the elliptically polarized mode. The relation $\Omega_{-\mathbf{k}} = \Omega_{+\mathbf{k}} \neq 0$ indicates the preserved spatial-inversion symmetry and broken time-reversal symmetry. And the $\Omega_{\mathbf{k}}$ in two-dimensional BZ shows that the three-fold rotational symmetry is broken, showing the gap opening from the mode with $1 > \tan{\eta} > 0$ is also related to anisotropy in a long-range distance (accompanied with three-fold rotational symmetry breaking). 

For zero ellipticity degree $\text{tan} \eta = 0$, i.e. a linearly polarized photon, the Berry curvature is zero everywhere in the BZ except on the flat-line band dispersion discussed in Sec.~\ref{sec:a_linearly_polarized_photon}, where its value can not be calculated since the wavefunctions  are singular. The zero Berry curvature for all the physical states indicates that the Dirac band gap induced by a linearly polarized mode is topologically trivial. 

\subsection{Two linearly polarized photons: isotropic case}
\label{sec:two_linearly_polarized_photons_isotropic}

\begin{figure}[!ht]
    \centering
    \includegraphics[width=1.0\linewidth]{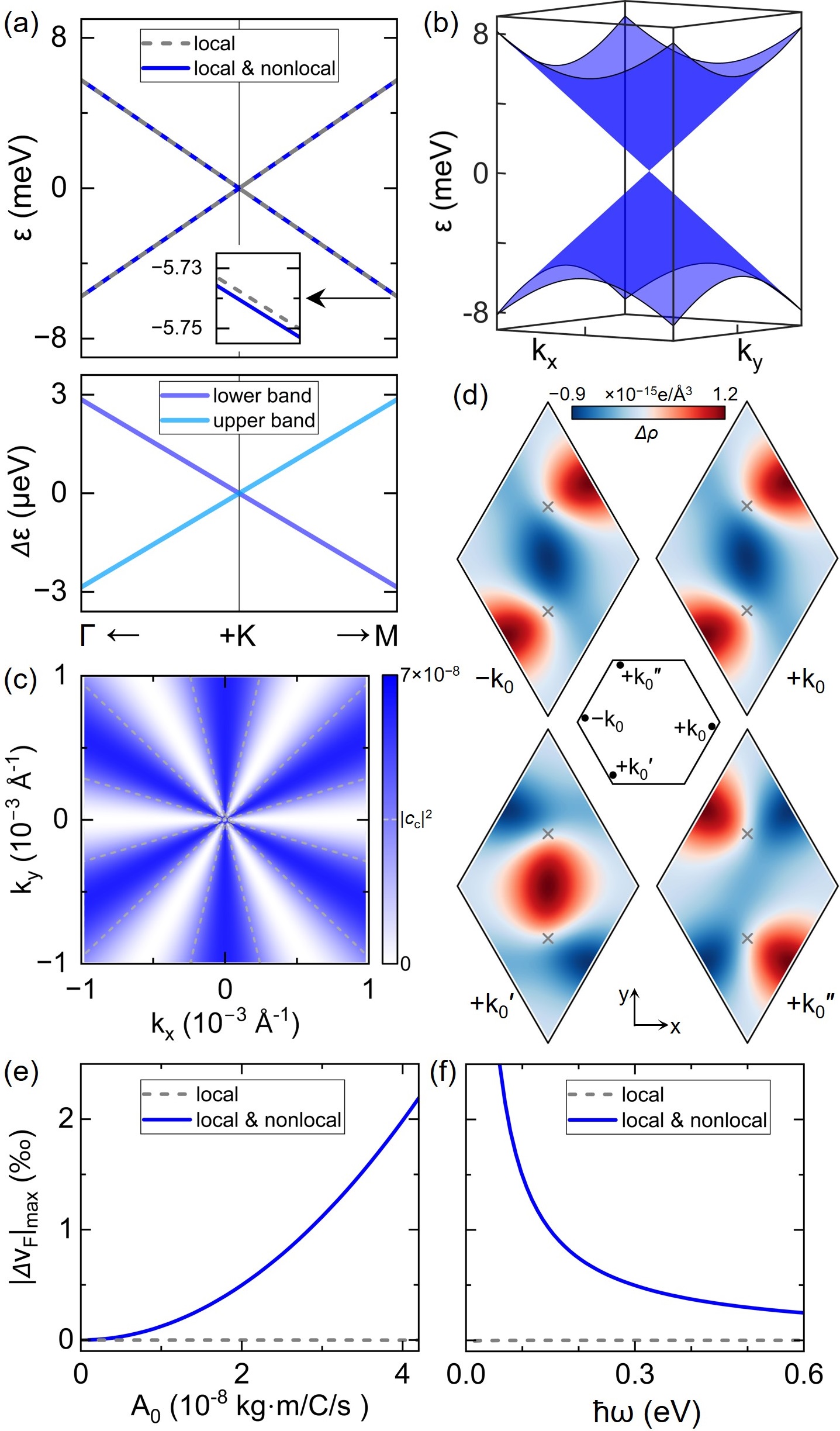}
    \caption{Dirac states in graphene coupled with two isotropic linearly polarized photon modes. The two modes have the same $\hbar\omega=0.3$ eV and $A_0 = 2 \times 10^{-8} \frac{\text{kg} \cdot \text{m}}{\text{C} \cdot \text{s}}$, but the perpendicular polarization vectors $\mathbf{e} = \mathbf{e}_x$ and $\mathbf{e}_y$. 
    (a) HF bands for the $+\mathbf{K}$ valley. The lower panel shows the variation of the bands due to addition of the nonlocal interaction. 
    (b) Representation of the blue band in (a) in $\{k_x , k_y\} \in [-1,1]$~$10^{-3}$\text{\AA}$^{-1}$ centered at $+\mathbf{K}$. 
    (c) Component $|c_c|^2$ of the conduction basis state $\varphi_{c\mathbf{k}}^0$ for the lower valence band ($\varphi_{v\mathbf{k}} = c_v \varphi_{v\mathbf{k}}^0 + c_c \varphi_{c\mathbf{k}}^0$) in (b). 
    (d) Variation of the electron density, $\Delta \rho (\mathbf{r}) = \sum_\mathbf{k} |\varphi_{v \mathbf{k}}|^2 - |\varphi_{v \mathbf{k}}^0|^2$, at $z = 0.33~\text{\AA}$ plane. The four panels show the contribution from the valence states with crystal momenta $-\mathbf{k}_0$, $+\mathbf{k}_0$, $+\mathbf{k}_0^\prime$ and $+\mathbf{k}_0^{\prime\prime}$. $-\mathbf{k}_0$ is inversely symmetric to $+\mathbf{k}_0$, and the $+\mathbf{k}_0^\prime$ and $+\mathbf{k}_0^{\prime\prime}$ are rotated by $120^{\circ}$ and $240^{\circ}$ with respect to  $+\mathbf{k}_0$.
    (e,f) Evolution of the Fermi velocity along $+\mathbf{K} \rightarrow \mathbf{M}$ as a function of $A_0$ (with fixed $\hbar\omega=0.3$ eV) and $\hbar \omega$ (with fixed $A_0 = 2 \times 10^{-8} \frac{\text{kg} \cdot \text{m}}{\text{C} \cdot \text{s}}$), respectively.}
    \label{fig:2linear}
\end{figure}

Experimentally, it might be challenging to set up a cavity with a perfectly linearly polarized single photon mode. Hence, for a more general description, we discuss the case of two photon modes linearly polarized along the planar directions $x$ and $y$. We start by considering an isotropic cavity with two degenerate linearly polarized photons with the same mode amplitude $A_0 = 2 \times 10^{-8} \frac{\text{kg} \cdot \text{m}}{\text{C} \cdot \text{s}}$ and the same photon energy $\hbar\omega = 0.3$ eV, as before. As shown in Fig. \ref{fig:2linear}(a,b), there is a very small difference (on the order of $\sim 1$ $\mu$eV) of energy eigenvlues in the renormalized Dirac bands when only local or both local and nonlocal interactions are included. Notably, the renormalized Dirac cones remain gapless, while the Fermi velocity is slightly modified. Along the $+\mathbf{K} \rightarrow \mathbf{M}$ paths, the Dirac Fermi velocity at the valley has the maximum variation of 0.5\textperthousand.

To show the effects of the two photons, we study again the wavefunctions and density of electrons for the renormalized Dirac cones. As shown in Fig. \ref{fig:2linear}(c), the original Dirac valence and conduction states hybridize with each other. Compared with that in the case of a circularly and linearly polarized photon, respectively, in Fig. \ref{fig:circular}(c) and \ref{fig:linear}(c), the states hybridization at the $+\mathbf{K}$ valley is significantly weaker in the case of the two isotropic linearly polarized photons. Correspondingly, the renormalization of Dirac bands in Fig. \ref{fig:2linear}(a,b) is much weaker than that in the case of a circularly and linearly polarized photon, shown in Fig. \ref{fig:circular}(a,b) and \ref{fig:linear}(a,b), respectively.

The modification of the electron density is analyzed for the four valence states corresponding to the generic crystal momenta $ -\mathbf{k}_0$, $+\mathbf{k}_0$, $+\mathbf{k}^\prime_0$, and $+\mathbf{k}^{\prime\prime}_0 $ at the $\pm \mathbf{K}$ valleys [inset in Fig. \ref{fig:2linear}(d)]. 
The momenta $-\mathbf{k}_0$ and $+\mathbf{k}_0$ are related by inversion symmetry, and $+\mathbf{k}_0, +\mathbf{k}^\prime_0$ and $ +\mathbf{k}^{\prime\prime}_0$ are connected by three-fold rotational symmetry. As shown in Fig. \ref{fig:2linear}(d), the density variation for each of the four crystal momenta satisfies $\Delta \rho (\mathbf{r}) = \Delta \rho (-\mathbf{r})$, showing that the interacting system has spatial-inversion symmetry. At the same time, $\Delta \rho_{-\mathbf{k}_0} (\mathbf{r})= \Delta \rho_{+\mathbf{k}_0} (\mathbf{r})$ holds, indicating that time-reversal symmetry is preserved. Furthermore, the density variation for $+\mathbf{k}_0$, $+\mathbf{k}^\prime_0$ and $+\mathbf{k}^{\prime\prime}_0$ satisfies $\Delta \rho_{+\mathbf{k}_0} (\mathbf{r}) = \Delta \rho_{+\mathbf{k}^{\prime}_0} (\hat{C}_3 \mathbf{r}) = \Delta \rho_{+\mathbf{k}^{\prime \prime}_0} (\hat{C}_3 \hat{C}_3 \mathbf{r})$, with $\hat{C}_3$ three-fold rotational operator, showing that the interacting system is three-fold rotation symmetric. Thus, the two isotropic linearly polarized photons do not break the symmetries of graphene, and consequently the Dirac cones in the electron-photon interacting system remain gapless, while the Fermi velocity is modified. In Appendix \ref{sec:appendix_analyticalHF}, the analytical QED-HF solutions also suggest that the Dirac cones remain gapless for graphene interacting with two isotropic linearly polarized photons. 

By changing the amplitude and frequency of the two linearly polarized photon modes, the renormalization of the Dirac Fermi velocity can be modified. As shown in Fig. \ref{fig:2linear}(e,f), for a fixed photon energy $\hbar \omega = 0.3$ eV, the Fermi velocity increases with the mode amplitude; with fixed mode amplitude $A_0 = 2 \times 10^{-8} \frac{\text{kg} \cdot \text{m}}{\text{C} \cdot \text{s}}$, instead, the Fermi velocity decreases with increasing frequency. This shows the the variation of the Dirac Fermi velocity follows the relation $|\Delta v_{\text{F}}|=\tau \frac{A_0^2}{\omega}$, where $\tau$ is determined by the momentum elements of graphene. If only local interaction is included, no changes in the Dirac Fermi velocity are observed.

\subsection{Two linearly polarized photons: anisotropic case}
\label{sec:two_linearly_polarized_photons_anisotropic}

We now study the evolution of the Dirac states in graphene when changing the two linearly polarized photon modes from isotropic to anisotropic. For the two modes, the photon energy is fixed to be the same as before; the amplitude of the $x$-mode is fixed as $A_{0x} = 2 \times 10^{-8} \frac{\text{kg} \cdot \text{m}}{\text{C} \cdot \text{s}}$, while the amplitude of the $y$-mode is tuned between $A_{0y} = 0$ to $A_{0y} = A_{0x}$. Such a cavity setup could be realized, for instance, in a cavity whose mirrors are made of a material with an anisotropic dielectric function. The the most convenient way to represent the two cavity photon modes is by aligning them along the principal axes, which are assumed to be the perpendicular $x$ and $y$ directions here. Any other choice would be equivalent, but would result in additional mode mixing terms in the Hamiltonian. As shown in Fig. \ref{fig:topology}(d), we observe a smooth transition from the completely anisotropic case, 
with a Dirac band gap $\Delta \sim 2 $ meV induced by a single linearly polarized photon (Fig. \ref{fig:linear}), 
to the completely isotropic case, with no Dirac band gap instead (Fig. \ref{fig:2linear}). We stress once again that this behavior holds only if the photon-induced nonlocal interaction is included. In Fig. \ref{fig:topology}(e), the bands along the $-\mathbf{K} \rightarrow +\mathbf{K}$ path are shown for different ratios $\frac{A_{0y}}{A_{0x}}$, evolving from the anisotropic to the isotropic case. Along with the reduction of gap size to zero, the length of the singular flat line of the Dirac bands (Fig. \ref{fig:linear}) gradually decreases to zero. 

As shown in Fig. \ref{fig:topology}(f), the Berry curvature $\Omega_{\mathbf{k}}$ of the valence Dirac band is zero and it is undefined for the crystal momenta $\pm \mathbf{K}$, which shows that the Dirac gap induced by the two linearly polarized photon modes is topologically trivial regardless of the anisotropy. Note that, also for the graphene outside the cavity, the $\Omega_{\mathbf{k}}$ is zero and undefined for the $\pm \mathbf{K}$. Additionally, the relation $\Omega_{-\mathbf{k}} = \Omega_{+\mathbf{k}} = 0$ indicates that both time-reversal and spatial-inversion symmetries are preserved. The narrowing of Dirac gap is related to the recovery of isotropy (together with three-fold rotational symmetry) when approaching the case of two isotropic photon modes, i.e., $\frac{A_{0y}}{A_{0x}} \rightarrow 1$. 

\subsection{Effective hopping integrals}
\label{sec:effective_hopping_integrals}

\begin{figure*}[tb]
    \centering
    \includegraphics[width=0.9 \linewidth]{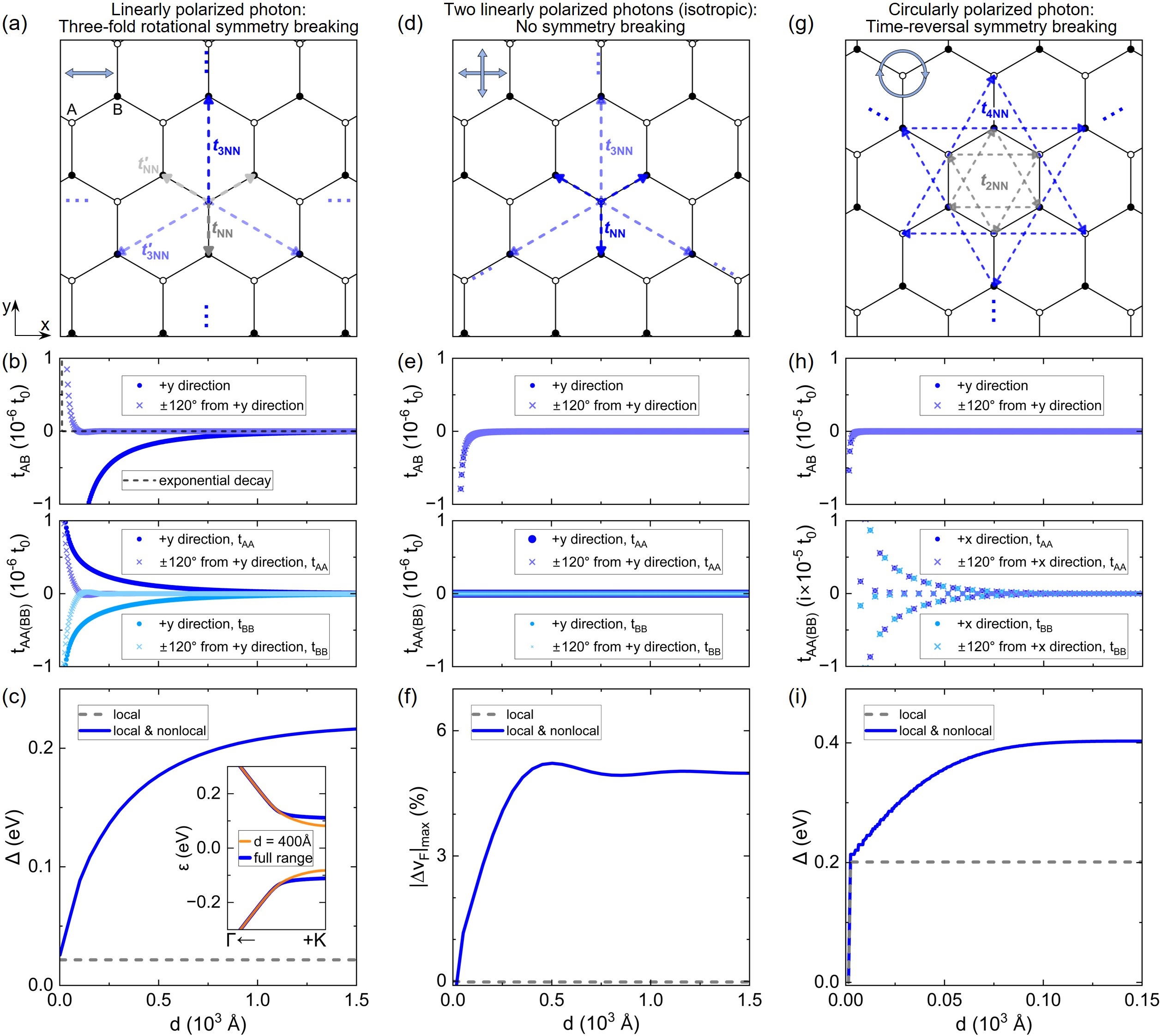}
    \caption{Cavity-mediated effective hopping integrals in graphene coupled to cavity photons of different polarizations. The photon energy is $\hbar\omega=0.3$ eV, and the mode amplitude is $A_0 = 2 \times 10^{-7} \frac{\text{kg} \cdot \text{m}}{\text{C} \cdot \text{s}}$ in all cases. 
    (a) Schematic of the effective electron hopping integrals from $A$ to $B$ sites (unfilled and filled dots, respectively), induced by a linearly polarized mode ($\mathbf{e}=\mathbf{e}_x$). The gray and light gray dashed lines show the NN integrals $t_\text{NN}$ and $t^{\prime}_\text{NN}$, while the blue and light blue dashed lines show the third NN integrals $t_\text{3NN}$ and $t^{\prime}_\text{3NN}$. The panel illustrates anisotropic hopping integrals, where the three-fold rotational symmetry in graphene is broken. The blue dots highlight that the effective hopping integrals from cavity-mediated nonlocal interactions are finite over a long range.
    (b) Hopping integrals $t_{AB}$, induced by a linearly polarized mode, change with the distance between sites for graphene (For numerical stability we use a non-significant sublattice potential $V_{AB} = \pm 0.004t_0$). The $t_{AB}$, $t_{AA}$ and $t_{BB}$ along the direction of $+y$ and the directions rotated by $\pm120^\circ$ from $+y$ are shown. The hopping integrals with exponential decay in intrinsic graphene are shown for comparison. 
    (c) The size of the cavity-induced Dirac band gap increases by including the effective hopping integrals in a longer range. 
    The inset shows the HF bands along $+\mathbf{K} \rightarrow \mathbf{\Gamma}$ for truncated and full interaction ranges. 
    (d) Schematic of the effective hopping integrals from $A$ to $B$ sites, induced by two isotropic linearly polarized modes with the same $\hbar\omega$ and $A_0$, and perpendicular polarization vectors $\mathbf{e}=\mathbf{e}_x$ and $\mathbf{e}_y$. The dashed blue lines represent the modified hopping integrals from $A$ to $B$ sites from nonlocal interaction, which does not break the symmetries in intrinsic graphene. 
    (e) The hopping integrals $t_{AB}$, induced by two isotropic linearly polarized modes, change with the distance between $A\text{-}B$ sites. The $t_{AA}$ and $t_{BB}$ remain unchanged at zero. 
    (f) The cavity-renormalized Dirac Fermi velocity changes by including hopping integrals over a longer range similar to the single linearly polarized mode. The renormalized Dirac states in the case of two isotropic linearly polarized modes remain gapless since the interacting system has the same symmetries as that in intrinsic graphene, as illustrated in (d). 
    (g) Schematic of the effective hopping integrals induced by a circularly polarized mode with $\mathbf{e} = \mathbf{e}_x + i\mathbf{e}_y$. The dashed gray lines show the second NN SOC-like integrals from local interaction, while the blue dashed lines show the long-range SOC integrals from nonlocal interaction. This indicates that time-reversal symmetry in graphene is broken by the effective SOC magnetic flux.
    (h) The hopping integrals, induced by a circularly polarized mode, change with the distance between sites. The integrals $t_{AB}$ along the $+y$ direction and the directions rotated by $\pm120^\circ$ from $+y$ are shown. The SOC integrals between $A\text{-}A$ (and $B\text{-}B$) sites, along the $+x$ direction and the direction rotated by $\pm120^\circ$ from $+x$, are shown. 
    (i) The size of the cavity-induced band gap quickly reaches a plateau value while increasing the range of the hopping, indicating a more localized interaction than that in the case of linear polarization in (c). 
    }
    \label{fig:hopping}
\end{figure*}

In Sec. \ref{sec:a_linearly_polarized_photon}, we suggested that the presence of the singular flat line in the band structure of graphene coupled to a single linearly polarized photon is a symptom of the unphysical infinite-range description of the photon-induced interactions. To corroborate this claim, we here analyze the electron interactions in a real space tight-binding representation, where we map the interactions (both local and nonlocal) to hopping terms and show how a physical description of the interaction can be recovered. We compute the mediated effective electron hopping integrals by photon-induced interactions for arbitrary range as follows: the QED Fock matrix ${F}_{\{v,c\}\,\mathbf{k}}$ in the basis set $\{ \varphi_{v \mathbf{k}}^0 , \varphi_{c \mathbf{k}}^0  \}$, directly obtained from the calculations in $\mathbf{k}$ space (as performed in all sections above), is transformed into ${F}_{\{A,B\}\,\mathbf{k}}$  
\begin{equation}
\label{eq:trans_FAB}
{F}_{\{A,B\}\,\mathbf{k}} = \begin{bmatrix} {F}_{AA\,\mathbf{k}} & {F}_{AB\,\mathbf{k}} \\ {F}_{BA\,\mathbf{k}} & {F}_{BB\,\mathbf{k}} \\ \end{bmatrix}
\end{equation}
in the $AB$-site representation with basis set $\{ \varphi_{A \mathbf{k}}^0 , \varphi_{B \mathbf{k}}^0 \}$, where $2p_z$ orbitals are used for the sites and the elements are constructed as $F_{IJ\,\mathbf{k}} = \sum_{\mathbf{R}} e^{i\mathbf{k} \cdot \mathbf{R}} t_{0I,\mathbf{R}J}$ with $I,J\in \{A,B\}$ and cell index $\mathbf{R}$ (see Appendix \ref{sec:appendix_grapheneTB} for the details of lattice structure). The effective electron hopping integrals in real space can be found from the Fourier transformation  $t_{0I,\mathbf{R}J} = \frac{1}{N_{\mathbf{k}}} \sum_{\mathbf{k}} e^{-i\mathbf{k} \cdot \mathbf{R}} F_{IJ\,\mathbf{k}}$. Finally, the range of the photon-induced interactions can be selected by truncating the effective hoppings in real space and performing an inverse Fourier transformation to get the QED Fock matrix in momentum space.

For the following results, we adopt the photon mode with energy of $\hbar \omega = 0.3$ eV and amplitude of $A_0 = 2 \times 10^{-7} \frac{\text{kg} \cdot \text{m}}{\text{C} \cdot \text{s}}$.  
The most interesting case
is that of graphene coupled to a single linearly polarized photon. To aid numerical convergence, we introduced a non-significant sublattice potential difference $V_{AB} = \pm 0.004t_0$ (which corresponds to a band gap of $0.022$ eV in the bare graphene); as shown later, the energy scales, affected by the range of the considered cavity-mediated interactions, are much larger than the one introduced by this sublattice potential. The schematic in Fig.~\ref{fig:hopping}(a) highlights the anisotropy at all orders of the hopping integrals by showing that the three-fold rotational symmetry of the original graphene is lost. 
Indeed, figure \ref{fig:hopping}(b) shows that the change in the electron hopping integrals, when including the cavity-induced nonlocal interaction, is finite and anisotropic 
over a long range of distance (over hundreds of nm, which is comparable to characteristic cavity confinement length \cite{Mark_theory}). The hoppings decay much slower with distance than the exponentially decaying ones in pristine graphene. In appendix \ref{sec:appendix_grapheneFock} and Sec. \ref{sec:a_linearly_polarized_photon}, we show that while $t_{AB}$ provide an increase of the distance between conduction and valence bands across the BZ, the modified sublattice hoppings $t_{AA}$ and $t_{BB}$ provide a further modification to the bands close to the $\pm \mathbf{K}$ valleys.  

Cutting the range of interaction (i.e. setting the effective hopping integrals to zero for a distance larger than a given threshold) has a direct effect on the size of the Dirac gap, as shown in Fig. \ref{fig:hopping}(c): long-range hopping integrals up to at least $\sim 150$ nm are needed to reach convergence on the gap size of $\sim 0.22$ eV (note that this value is significantly larger than that from the artificial sublattice potential $V_{AB}$ used for numerical stability). Crucially, the inset in Fig.~\ref{fig:hopping}(c) demonstrates that the kinks on the flat line in bands and the singular behavior of wavefunctions, discussed in Sec.~\ref{sec:a_linearly_polarized_photon}, are removed when the interaction range is truncated. This confirms that the singularity is a consequence of the infinite range of the photon-induced interactions which no realistic cavities could support. 

For comparison, we performed an equivalent analysis for the case of two isotropic linearly polarized photons and the case of a single circularly polarized photon. 
In the presence of two isotropic linearly polarized cavity photon modes [Fig. \ref{fig:hopping}(d,e)], we observe a range of interaction similar to the case for a single linearly polarized photon, but as discussed in Sec. ~\ref{sec:two_linearly_polarized_photons_isotropic}, because no symmetry is broken, no electronic Dirac gap is opened. Nevertheless, the renormalization of the Fermi velocity at the $\pm \mathbf{K}$ valleys shows a dependence on the interaction range [Fig. \ref{fig:hopping}(f)].
Finally, for a circularly polarized photon, Fig.~\ref{fig:hopping}(g) illustrates that spin-orbit-coupling (SOC) like effective electron hoppings are induced (already at the level of cavity-induced local interactions), indicating time-reversal symmetry is broken. Effectively, the interacting electron-photon system can be mapped to an Haldane model. The cavity-induced nonlocal interaction introduces long-range effective SOC electron hoppings [Fig. \ref{fig:hopping}(h,i)], up to a distance of $\sim 10$ nm which is considerably smaller than that in Fig. \ref{fig:hopping}(b,c) for a single linearly polarized mode.

\section{Conclusions}
In conclusion, a non-perturbative QED-HF theoretical approach based on the photon-free self-consistent framework is formulated to address the photon-mediated local and nonlocal interactions of electrons in the extended materials collectively coupled with the fluctuating photons in optical cavities. The photon-induced nonlocal electron-electron interaction, originating from the quantum nature of light confined by a cavity, naturally arises in the formulation of our photon-free QED-HF approach. Such a nonlocal interaction, quite different from the Coulomb charge-charge repulsion interaction, is properly treated through the self-consistent iterations of electronic wavefunctions in the coupling systems of crystalline matter and optical cavities. 

Using the QED-HF approach, we have shown that photon-induced nonlocal interaction plays a significant role in the renormalization of the Dirac electronic properties in graphene. By accounting for only the photon-mediated local interaction, a circularly polarized photon opens a topologically nontrivial Dirac gap from time-reversal symmetry breaking, while a linearly polarized photon does not lift the band degeneracy at the Dirac points. In contrast, with the photon-induced nonlocal electron-electron interaction included, both the circularly and linearly polarized photon modes open a Dirac band gap, with nontrivial and trivial electronic topology, respectively. The formation of topologically trivial gap with flat-band dispersion is related to the anisotropy in the presence of photon-induced interaction over a long range. With two isotropic linearly polarized photons, all the symmetries in intrinsic graphene are restored and the Dirac cones remain gapless while the Dirac Fermi velocity is slightly affected purely by the photon-induced nonlocal interaction. Importantly, we repeatedly show that the non-perturbative nature of the QED-HF approach is crucial for describing the renormalization of the Dirac electronic structure by cavity photons, demonstrating that the non-perturbative treatment is a necessity to capture the key quantum features (missed in perturbative methods, even qualitatively) resulted from the collectivity of electron-photon coupling inherent to the quantum nature of cavity photons. 

As a general theoretical framework based on wavefunction iterations, our photon-free QED-HF approach can be employed to study a wide range of electron-photon interacting material systems that inherently possess the photon-mediated electron-electron interaction over large distances due to quantum cavity confinement effects. This opens the door to predicting novel cavity-induced quantum phenomena in extended crystalline matter, and providing deeper insights into how quantum confinement effects mediate long-range interactions and fundamentally influence material properties. Also, the photon-free QED-HF method can be integrated into \textit{ab-initio} first-principles modeling packages to investigate the realistic materials coupled with cavity photon modes, where the collective coupling between ions and photons can also be addressed within the QED-HF framework by including the phononic degrees of freedom.

\begin{acknowledgments}
 We thank Michael Ruggenthaler and I-Te Lu for the helpful discussions. We acknowledge support from the Cluster of Excellence ``CUI: Advanced Imaging of Matter"- EXC 2056 - project ID 390715994, SFB-925  of the Deutsche Forschungsgemeinschaft (DFG), ERC  grant  and Grupos Consolidados (IT1453-22), and the Max Planck-New York City Center for Non-Equilibrium Quantum Phenomena. The Flatiron Institute is a division of the Simons Foundation. We acknowledge support from the European Union Marie Sklodowska-Curie Doctoral Networks TIMES grant No. 101118915 and SPARKLE grant No. 101169225. Hang Liu thanks the Alexander von Humboldt Foundation for the support from Humboldt Research Fellowship.  
\end{acknowledgments}

\appendix

\section{Mode transformation}
\label{sec:appendix_modetrans}
When considering a single photon mode, the operator of the electromagnetic vector potential is $\hat{\mathbf{A}}=A_0\left(\hat{a}^{\dagger}\mathbf{e}^*+\hat{a}\mathbf{e}\right)$. As shown in Eq. \eqref{eq:QED_Hamiltonian}, the interaction part of the Pauli-Fierz Hamiltonian contains a paramagnetic term with $\sum_i^{N_\text{e}}\hat{\mathbf{p}}_i \cdot \hat{\mathbf{A}}$ and a diamagnetic term with $\hat{\mathbf{A}}^2$. To simplify the Hamiltonian, the photon operators can be redefined to effectively remove the diamagnetic term. For a linearly polarized photon mode, since $\mathbf{e}^2 = \mathbf{e} \cdot \mathbf{e}^* = 1$, one has that $\hat{\mathbf{A}}^2 = A_0^2\left(\hat{a}^{\dagger2}+\hat{a}^2+2\hat{a}^{\dagger}\hat{a}+1\right)$, where $\hat{a}^{\dagger 2}$ and $\hat{a}^2$ can be eliminated to absorb the diamagnetic term into the bare photon term. To do this, a new photon operator $\hat{\tilde{a}}$ is introduced via the transformation  
$\left(\begin{array}{ll}
\hat{a} \\
\hat{a}^\dagger \\
\end{array}\right)=\left(\begin{array}{ll}
c_{1} & c_{2} \\
c_{2} & c_{1}
\end{array}\right)\left(\begin{array}{l}
\hat{\tilde{a}} \\
\hat{\tilde{a}}^{\dagger} \\
\end{array}\right)$ with coefficients $c_{1}=\sqrt{\frac{u + 1}{2}}$ and $c_{2}=-\sqrt{\frac{u - 1}{2}}$. With the dressed photon operators $\hat{\tilde{a}}$ and $\hat{\tilde{a}}^\dagger$, the sum of the diamagnetic term and the bare photon term is 
\begin{equation}
    \hat{H}_\text{dia} + \hat{H}_\text{p} = \hbar \omega \sqrt{1 + \zeta \frac{2N_\text{e} A_0^2}{\omega}}\left( \frac{1}{2} + \hat{\tilde{a}}^{\dagger} \hat{\tilde{a}} \right),
\end{equation}
and the paramagnetic term is
\begin{equation}
    \hat{H}_\text{para}=-\frac{q}{m} \sum_i^{N_\text{e}} \hat{\mathbf{p}}_i \cdot A_0 (c_1+c_2) \left(\hat{\tilde{a}}^{\dagger}\mathbf{e}^*+\hat{\tilde{a}} \mathbf{e}\right).
\end{equation}
Thus, the Hamiltonian for a linearly polarized photon mode becomes 
\begin{equation}
    \hat{H} = \hat{H}_\text{e} + \hbar \tilde{\omega} \left( \frac{1}{2} +\hat{\tilde{a}}^{\dagger} \hat{\tilde{a}} \right) - \frac{q}{m} \sum_i^{N_\text{e}} \hat{\mathbf{p}}_i \cdot \hat{\tilde{\mathbf{A}}},
\end{equation}
where the amplitude of the dressed vector-potential operator $\hat{\tilde{\mathbf{A}}} = \tilde{A}_0 \left(\hat{\tilde{a}}^{\dagger}\mathbf{e}^*+\hat{\tilde{a}} \mathbf{e}\right)$ is $\tilde{A}_0 = A_0 (c_1 + c_2)$, and the dressed photon frequency is $\tilde{\omega} = \omega \sqrt{1+\zeta\frac{2N_\text{e} A_0^2}{\omega}} $. This indicates that the diamagnetic term merely renormalizes the mode amplitude and frequency to effective ones. 

Differently, for a circularly polarized photon mode (e.g., with $\mathbf{e} = \mathbf{e}_x + i\mathbf{e}_y$), since $\mathbf{e}^2=0$ and $ \mathbf{e} \cdot \mathbf{e}^* = 1$, the square of vector-potential operator, $\hat{\mathbf{A}}^2 = A_0^2 \left( 2\hat{a}^{\dagger}\hat{a} + 1 \right)$, does not have the terms with $\hat{a}^2$ and $\hat{a}^{\dagger 2}$. The sum of the diamagnetic term and the bare photon term is 
\begin{equation}
    \hat{H}_\text{dia} + \hat{H}_\text{p} = \hbar \omega \left( 1 + \zeta \frac{N_\text{e} A_0^2}{\omega} \right) \left( \frac{1}{2} +\hat{a}^\dagger \hat{a} \right). 
\end{equation}
The photon operator $\hat{\tilde{a}} = \hat{a}$ with a dressed frequency $\tilde{\omega} = \omega \left( 1 + \zeta \frac{N_\text{e} A_0^2}{\omega} \right)$ and amplitude $\tilde{A}_0 = A_0$ is introduced. With the dressed photon operator, the Hamiltonian for a circularly polarized photon mode is 
\begin{equation}
    \hat{H} = \hat{H}_\text{e} + \hbar \tilde{\omega} \left( \frac{1}{2} +\hat{\tilde{a}}^{\dagger} \hat{\tilde{a}} \right) - \frac{q}{m} \sum_i^{N_\text{e}} \hat{\mathbf{p}}_i \cdot \hat{\tilde{\mathbf{A}}},
\end{equation}
where the dressed vector-potential operator is $\hat{\tilde{\mathbf{A}}} = \tilde{A}_0 \left(\hat{\tilde{a}}^{\dagger}\mathbf{e}^*+\hat{\tilde{a}} \mathbf{e}\right)$. Thus, for a circularly polarized photon mode, the diamagnetic terms only renormalize the photon frequency.

\section{Downfolding of QED Hamiltonian in a dressed photon space}
\label{sec:appendix_downfolding}
In the limit of high dressed photon frequency, the Hamiltonian in Eq. \eqref{eq:dressed_QED_Hamiltonian} and \eqref{eq:qed_hamiltonian_multimodes} can be effectively downfolded into the zero dressed photon sector \cite{Simone_PNAS,PhysRevLett.126.153603}. For the Hamiltonian in Eq. \eqref{eq:dressed_QED_Hamiltonian} with a single photon mode, the downfolding formula up to the first-order expansion is 
\begin{equation}
\hat{H}_\text{eff} = \langle \tilde{0} | \hat{H} | \tilde{0} \rangle - \frac{ \langle \tilde{0} | \hat{H} | \tilde{1} \rangle   \langle \tilde{1} | \hat{H} | \tilde{0} \rangle }{\hbar \tilde{\omega}} , 
\end{equation}
where the integrals are calculated as $\langle \tilde{0} | \hat{H} | \tilde{0} \rangle = \hat{H}_\text{e} + \frac{\hbar \tilde{\omega}}{2}$, $\langle \tilde{0} | \hat{H} | \tilde{1} \rangle = - \frac{q \tilde{A}_0}{m}  \sum_i^{N_\text{e}} \hat{\mathbf{p}}_i \cdot \mathbf{e}$, and $\langle \tilde{1} | \hat{H} | \tilde{0} \rangle = - \frac{q \tilde{A}_0}{m}  \sum_i^{N_\text{e}} \hat{\mathbf{p}}_i \cdot \mathbf{e}^* $. The downfolding gives rise to the photon-free QED Hamiltonian shown in Eq. \eqref{eq:effective_hamiltonian}. 
For the Hamiltonian in Eq. \eqref{eq:qed_hamiltonian_multimodes} with $N_\text{p}$ modes, the downfolding formula up to the first-order expansion is 
\begin{equation}
    \begin{aligned}
        \hat{H}_{\text{eff}} & =  \langle \tilde{0}_1, \cdots, \tilde{0}_{N_\text{p}} | \hat{H} | \tilde{0}_1, \cdots, \tilde{0}_{N_\text{p}} \rangle - \\ &  
         \sum_\alpha^{N_\text{p}} \bigg[   \langle \tilde{0}_1, \cdots,  \tilde{0}_{N_\text{p}} | \hat{H} | \tilde{0}_1, \cdots, \tilde{1}_\alpha, \cdots, \tilde{0}_{N_\text{p}} \rangle     \\ &   \cdot \langle \tilde{0}_1, \cdots, \tilde{1}_\alpha, \cdots, \tilde{0}_{N_\text{p}} | \hat{H} | \tilde{0}_1, \cdots, \tilde{0}_{N_\text{p}} \rangle \frac{1}{\hbar \tilde{\omega}_\alpha} \bigg]  
    \end{aligned},
\end{equation}
which is calculated to be the photon-free QED Hamiltonian in Eq. \eqref{eq:multimode_effective_hamiltonian}.

\section{Derivation of photon-free QED-HF equation}
\label{sec:appendix_HF}
Within the HF approximation, the wavefunction of the effective photon-free Hamiltonian in Eq. \eqref{eq:effective_hamiltonian} is represented by a single Slater determinant $\Psi(\mathbf{r}_1,\cdots,\mathbf{r}_{N_\text{e}}) = \text{det} \{\varphi_1, \cdots, \varphi_{N_\text{e}} \}$,
where $\{\varphi_i\}$ are the single-particle orbitals occupied by electrons, and the electronic spin degrees of freedom is not considered. The total energy of the many-electron system with photon-induced interactions is (see Appendix \ref{sec:appendix_Brillouin} for the Slater-Condon rules for the operators from photon-induced electron interactions) 
\begin{equation}
    \label{eq:total_energy_many_particle}
    \begin{aligned}
    E & = \langle \Psi  |\hat{H}_\text{eff} | \Psi  \rangle   = \frac{\hbar \tilde{\omega}}{2} + \sum_i^{N_\text{e}} \bigg[ \langle \varphi_i | \hat{h}_\text{e} + \hat{h}_\text{l}  | \varphi_i\rangle \\ 
    & + \sum_{j}^{N_\text{e}} \langle\varphi_i \varphi_j | \hat{h}_\text{nl}  | \varphi_i \varphi_j \rangle     - \sum_{j}^{N_\text{e}}  \langle\varphi_i \varphi_j | \hat{h}_\text{nl}  | \varphi_j \varphi_i \rangle \bigg]   
    \end{aligned}
\end{equation}
with 
\begin{equation}
    \hat{h}_\text{l} = - \zeta \frac{\tilde{A}_0^2}{\tilde{\omega}} \left( \hat{\mathbf{p}}_{\mathbf{r}} \cdot \mathbf{e} \right)\left( \hat{\mathbf{p}}_{\mathbf{r}} \cdot \mathbf{e}^* \right),
\end{equation}
and 
\begin{equation}
    \hat{h}_\text{nl} = - \zeta \frac{\tilde{A}_0^2}{\tilde{\omega}} \left( \hat{\mathbf{p}}_{\mathbf{r}} \cdot \mathbf{e} \right)\left( \hat{\mathbf{p}}_{\mathbf{r}^\prime} \cdot \mathbf{e}^* \right).
\end{equation}
The total energy is contributed by local, direct, and exchange interactions, and a constant dressed zero-point energy. Importantly, when taken together, the direct and exchange terms are self-interaction free since the self-direct and self-exchange interactions exactly cancel with each other. In Eq. \eqref{eq:total_energy_many_particle}, the single integral is for $\mathbf{r}$ coordinate, and the double integral is for $\mathbf{r}$ and $\mathbf{r}^\prime$, where the coordinates for the four orbital functions are $\mathbf{r}$, $\mathbf{r}^\prime$, $\mathbf{r}$, and $\mathbf{r}^\prime$ in sequence. The convention of coordinates for double integral is used in this article. 
To find the ground state, one can use the variational principle. Hence, we minimize the functional 
\begin{equation}
    G(\{\varphi_i\}) = E - \sum_i^{N_\text{e}}\varepsilon_i\left\langle\varphi_i | \varphi_i\right\rangle
\end{equation}
by varying the occupied orbitals. 

The first-order variation $\delta G$, arising from an infinitesimal variation of the occupied states $\varphi_i+\delta \varphi_i$, is contributed by photon-induced local and nonlocal interactions. 
Taking $\delta\varphi_i  =0$ and arbitrary $ \delta\varphi_i^*$, the component from the local single-electron interaction terms is
\begin{equation}
    \begin{aligned} 
        \delta G_1 = \sum_i^{N_\text{e}} \bigg[ \langle \delta \varphi_i | \hat{h}_\text{e}  + \hat{h}_\text{l}  | \varphi_i \rangle - \varepsilon_i\left\langle \delta \varphi_i | \varphi_i \right\rangle \bigg]  
    \end{aligned},
\end{equation}
and the component from the nonlocal electron-electron interaction terms is
\begin{equation}
    \begin{aligned} 
        \delta G_2 = & \sum_{i}^{N_\text{e}} \sum_{j}^{N_\text{e}}  \bigg[ \langle \delta \varphi_i  \varphi_j  |  \hat{h}_\text{nl}  +  \hat{h}_\text{nl}^\dagger      | \varphi_i  \varphi_j  \rangle  \\
        &        -    \langle \delta \varphi_i  \varphi_j  |  \hat{h}_\text{nl} +  \hat{h}_\text{nl}^\dagger | \varphi_j  \varphi_i  \rangle   \bigg] 
    \end{aligned}.
\end{equation}
For the total first-order functional variation $\delta G = \delta G_1 + \delta G_2$ to be zero for any choice of $\langle \delta\varphi_i |$, the QED-HF equation, given by Eq. \eqref{eq:hf_equation}-\eqref{eq:nonlocal_exchange}, must hold. The occupied orbitals obtained from the QED-HF equation define the ground-state determinant wavefunction $\Psi_{\text{GS}}$ with the lowest total energy $E_\text{GS}$.

\section{\textit{k} decoupling of QED Fock matrix}
\label{sec:appendix_Fock}
For a crystal, the Bloch state for a specific band $n$ and crystal momentum $\mathbf{k}$ is $\varphi_{n \mathbf{k}}^0 (\mathbf{r}) = e^{i \mathbf{k} \cdot \mathbf{r}} u_{n \mathbf{k}}^0 (\mathbf{r})$, where $u_{n \mathbf{k}}^0 (\mathbf{r}) = u_{n \mathbf{k}}^0 (\mathbf{r} + \mathbf{R})$ is a periodic function with period $\mathbf{R}$ of lattice in real space. The momentum matrix element between two Bloch states is  
\begin{equation}
\begin{aligned} 
    \langle \varphi_{n \mathbf{k}}^{0} | \hat{\mathbf{p}} | \varphi_{n^\prime \mathbf{k}^\prime}^{0} \rangle = \delta_{\mathbf{k}\mathbf{k}^\prime} \big[ \hbar \mathbf{k}  \delta_{nn^\prime}  - i \hbar  \langle u_{n \mathbf{k}}^{0*}  | \nabla | u_{n^\prime \mathbf{k} }^{0}  \rangle \big],
\end{aligned}    
\end{equation}
which is zero for any two Bloch states with different crystal momenta. 
As a result, with the Bloch states $\{ \varphi_{n\mathbf{k}}^0 \}$ without interaction to cavity photons as a basis set, the element of QED Fock matrix would be 
\begin{equation}
    F_{n\mathbf{k}~n^\prime \mathbf{k}^\prime} = \langle \varphi_{n \mathbf{k}}^{0}  | \hat{\mathcal{F}} | \varphi_{n^\prime \mathbf{k}^\prime}^{0}  \rangle = \delta_{\mathbf{k}\mathbf{k}^\prime} \langle \varphi_{n \mathbf{k}}^{0}  | \hat{\mathcal{F}}  | \varphi_{n^\prime \mathbf{k}}^{0}  \rangle, 
\end{equation}
which shows that the Fock matrix can be written out separately for each crystal momentum $\mathbf{k}$. Thus, in the implementation of numerical calculations, the QED Fock matrix for photon-induced interactions can be constructed and diagonalized for each crystal momentum individually. 
When the total momentum $\mathbf{P} = 0$, the direct component in Eq. \eqref{eq:matrixdirect} vanishes, meaning that the construction of the Fock matrix for a given crystal momentum $\mathbf{k}$ involves only the states with the same $\mathbf{k}$, as discussed in Sec. \ref{sec:QED_HF}. This avoids summing over all crystal momenta $\mathbf{k}^\prime$ in the QED-HF formulation. In numerical calculations, enforcing the zero total momentum (and thus a zero direct component) is not necessary, as the direct component can be explicitly computed.

To initiate self-consistent calculations on the QED Fock matrix, the electronic Bloch states $\{ \varphi_{n\mathbf{k}}^0 \}$ or their linear combinations can be used to construct the initial HF potential. 
The resulting Fock matrix is then solved, and the HF potential and wavefunctions are iteratively updated until convergence is reached, where the convergence can be monitored by setting a threshold for the HF eigenvalue, potential, or total energy.

\section{Slater-Condon rules and Brillouin's theorem for photon-induced electron interactions}
\label{sec:appendix_Brillouin}
The Slater determinant for $N_\text{e}$ electrons $\Psi = \text{det}\{\varphi_1, \cdots, \varphi_i, \cdots, \varphi_j, \cdots, \varphi_{N_\text{e}} \}$ is constructed from $N_\text{e}$ single-particle orbitals occupied by electrons. With the excitation of an electron from orbital $\varphi_i$ to $\varphi_k$, $\Psi$ is singly excited and becomes
$\Psi_i^k = \text{det}\{\varphi_1, \cdots, \varphi_k, \cdots, \varphi_j, \cdots, \varphi_{N_\text{e}} \}$; with the excitation of two electrons from orbitals $\varphi_i$ and $\varphi_j$ to $\varphi_k$ and $\varphi_l$, $\Psi$ is doubly excited to $\Psi_{ij}^{kl} = \text{det}\{\varphi_1, \cdots, \varphi_k, \cdots, \varphi_l, \cdots, \varphi_{N_\text{e}} \}$. The overlap integrals of the photon-induced local operator $\hat{H}_\text{l}$ in Eq. \eqref{eq:effective_hamiltonian_localpart} between two determinants are 
\begin{equation}
\label{eq:one_body_slater_condon1}
\langle\Psi | \hat{H}_\text{l}| \Psi  \rangle = \sum_i^{N_\text{e}} \langle \varphi_i | \hat{h}_\text{l} | \varphi_i \rangle,
\end{equation}
and
\begin{equation}
\label{eq:one_body_slater_condon2}
\langle\Psi | \hat{H}_\text{l} | \Psi_i^k \rangle =  \langle \varphi_i | \hat{h}_\text{l} | \varphi_k \rangle.
\end{equation}
For the two determinants differing by two or more orbitals, the overlap integral of $\hat{H}_\text{l}$ is zero. 

The overlap integrals of the photon-induced nonlocal operator $\hat{H}_\text{nl}$ in Eq. \eqref{eq:effective_hamiltonian_nonlocalpart} between two determinants are
\begin{equation}
\label{eq:two_body_slater_condon1}
\langle \Psi | \hat{H}_\text{nl}| \Psi  \rangle=  \sum_i^{N_\text{e}} \sum_{j \neq i}^{N_\text{e}}  \langle\varphi_i \varphi_j|\hat{h}_\text{nl}| \varphi_i \varphi_j \rangle- \langle\varphi_i \varphi_j|\hat{h}_\text{nl}| \varphi_j \varphi_i \rangle,  
\end{equation}
\begin{equation}
\label{eq:two_body_slater_condon2}
\begin{aligned}
& \langle\Psi |\hat{H}_\text{nl}| \Psi_i^k \rangle=  \sum_j^{N_\text{e}}   \langle\varphi_j \varphi_i|\hat{h}_\text{nl}| \varphi_j \varphi_k \rangle -  \langle\varphi_j \varphi_i|\hat{h}_\text{nl}| \varphi_k \varphi_j \rangle  \\
& \quad \quad \quad \quad \quad \quad ~    +  \langle\varphi_i \varphi_j|\hat{h}_\text{nl}| \varphi_k \varphi_j \rangle- \langle\varphi_i \varphi_j|\hat{h}_\text{nl}| \varphi_j \varphi_k \rangle  
\end{aligned},
\end{equation}
and
\begin{equation}
\label{eq:two_body_slater_condon3}
\begin{aligned}
& \langle\Psi |\hat{H}_\text{nl}| \Psi_{i j}^{k l} \rangle=   \langle\varphi_i \varphi_j |\hat{h}_\text{nl}| \varphi_k \varphi_l \rangle -  \langle\varphi_i \varphi_j|\hat{h}_\text{nl}| \varphi_l \varphi_k \rangle \\
& \quad \quad \quad \quad \quad ~  +   \langle\varphi_j \varphi_i |\hat{h}_\text{nl}| \varphi_l \varphi_k \rangle- \langle\varphi_j \varphi_i |\hat{h}_\text{nl}| \varphi_k \varphi_l \rangle
\end{aligned}.
\end{equation}
For the two determinants with three or more different orbitals, the overlap integral of $\hat{H}_\text{nl}$ is zero. Equations \eqref{eq:one_body_slater_condon1}-\eqref{eq:two_body_slater_condon3} are the Slater-Condon rules for the operators from photon-induced electron interactions. 

The Slater-Condon rules in Eq. \eqref{eq:two_body_slater_condon1}-\eqref{eq:two_body_slater_condon3} for photon-induced nonlocal electron-electron operator can be compared with that for the Coulomb electron-electron operator. The latter does not change if the electron indexes $i$ and $j$ are exchanged $\frac{1}{|\mathbf{r}_i-\mathbf{r}_j|}=\frac{1}{|\mathbf{r}_j-\mathbf{r}_i|}$, whereas the photon-induced operator would be changed with exchanged electron indexes $\left( \hat{\mathbf{p}}_i \cdot \mathbf{e}\right) \left( \hat{\mathbf{p}}_j \cdot \mathbf{e}^* \right) \neq \left( \hat{\mathbf{p}}_j \cdot \mathbf{e}\right) \left( \hat{\mathbf{p}}_i \cdot \mathbf{e}^* \right)$ for the circularly and elliptically polarized cavity modes due to $\mathbf{e} \neq \mathbf{e}^*$. Thus, the rules in Eq. \eqref{eq:two_body_slater_condon1}-\eqref{eq:two_body_slater_condon3} with circular or elliptical polarization differ from the well-known Slater-Condon rules for the Coulomb electron-electron operator. In contrast, if the cavity photon mode is linearly polarized, due to $\mathbf{e} = \mathbf{e}^*$, the exchange of electron indexes $i$ and $j$ does not change the photon-induced operator, thus the rules shown in Eq. \eqref{eq:two_body_slater_condon1}-\eqref{eq:two_body_slater_condon3}  for linearly polarized photon modes would be the same as that for the Coulomb electron-electron operator. 

Based on the Slater-Condon rules for photon-induced electron interactions, the matrix element of photon-free Hamiltonian in Eq. \eqref{eq:effective_hamiltonian} between the ground state determinant $\Psi_\text{GS}$ and its singly-excited determinant $\Psi_i^k$ is
\begin{equation}
    \begin{aligned}
        \langle\Psi_\text{GS}|\hat{H}_\text{eff}|\Psi_i^k\rangle  = \langle \varphi_i  | \hat{\mathcal{F}}  | \varphi_k  \rangle. 
    \end{aligned}
\end{equation}
When the determinants are constructed from QED-HF orbitals, the following is automatically satisfied
\begin{equation}
        \begin{aligned}    
        \langle\Psi_\text{GS}^\text{HF}|\hat{H}_\text{eff}|\Psi_i^{k~\text{HF}}\rangle = \langle \varphi_i^\text{HF}  | \hat{\mathcal{F}}  | \varphi_k^\text{HF}  \rangle = 0, 
    \end{aligned}
\end{equation}
which shows that the photon-free QED-HF approximation accounts for the interaction between the ground and first-excited configurations. This is equivalent to the Brillouin's theorem for electron-electron Coulomb interaction.

\section{Tight-binding model of graphene and momentum matrix elements}
\label{sec:appendix_grapheneTB}
\begin{figure}[!t]
    \centering
    \includegraphics[width=1.0\linewidth]{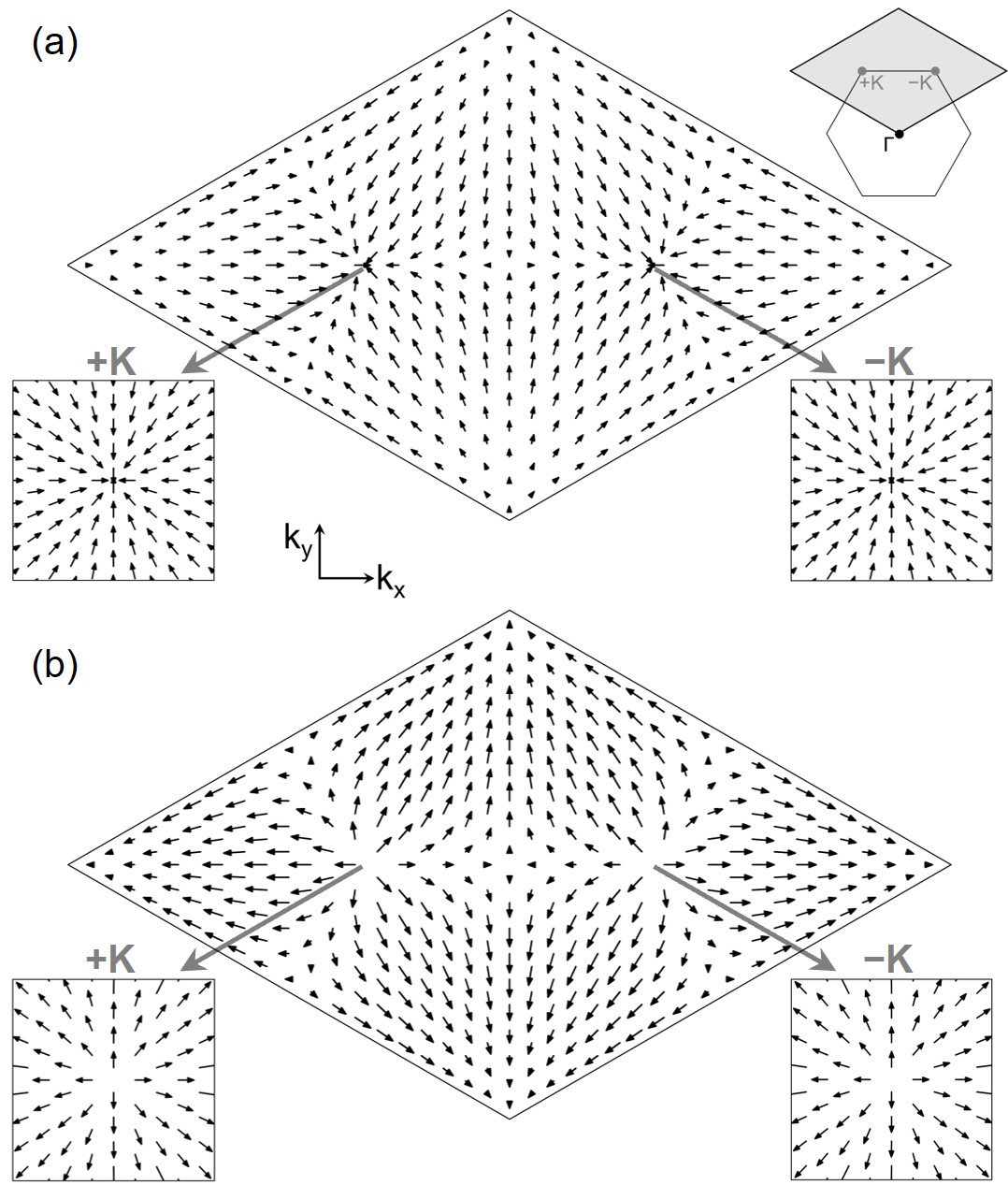}
    \caption{Electron momentum $\mathbf{p}_{vv\mathbf{k}}$ of valence states (a) and  $\mathbf{p}_{cc\mathbf{k}}$ of conduction states (b) in the full BZ of graphene. The length and orientation of arrows represent the amplitude and direction of electron momentum, respectively. The four square insets show the electron momentum of the Bloch states at the $\pm \mathbf{K}$ valleys. The shaded area in the upright inset of (a) illustrates the full BZ.}
    \label{fig:momentum}
\end{figure}

A two-band tight-binding model in the full BZ is used to compute the electronic states of monolayer graphene. The lattice vectors are set to $\mathbf{a}_{1}=(\frac{1}{2},\frac{\sqrt{3}}{2})a$ and $\mathbf{a}_{2}=(-\frac{1}{2},\frac{\sqrt{3}}{2})a$ with lattice constant $a=2.46$~\AA. In each unit cell, there are two lattice sites, labeled as $A$ and $B$, where a $2p_z$ orbital sits on each site. With crystal periodicity $\mathbf{R} = n_{\mathbf{a}_1} \mathbf{a}_1 + n_{\mathbf{a}_2} \mathbf{a}_2$ ($n_{\mathbf{a}_1},n_{\mathbf{a}_2} \in \mathbb{Z}$), the spatial coordinates of $A$ and $B$ sites in the whole space are $\mathbf{r}_{A} = \frac{2}{3}(\mathbf{a}_1 + \mathbf{a}_2) + \mathbf{R}$ and $\mathbf{r}_{B} = \frac{1}{3}(\mathbf{a}_1 + \mathbf{a}_2) + \mathbf{R}$, respectively. Using the $AB$-site Bloch functions $\big\{ \varphi_{A\mathbf{k}}^0 = \frac{1}{\sqrt{N_\text{cell}}} \sum_\mathbf{R}^{N_\text{cell}} e^{i \mathbf{k} \cdot \mathbf{R}} \varphi_{2p_z}(\mathbf{r}-\mathbf{r}_{A}),~\varphi_{B\mathbf{k}}^0 = \frac{1}{\sqrt{N_\text{cell}}} \sum_\mathbf{R}^{N_\text{cell}} e^{i \mathbf{k} \cdot \mathbf{R}} \varphi_{2p_z}(\mathbf{r}-\mathbf{r}_{B}) \big\}$ ($N_\text{cell}$ is the number of cells)  as a basis set, the Hamiltonian matrix of graphene with electron hopping energies between $A$ and $B$ sites is 
\begin{equation}
    \label{eq:graphene_h}
    h_\text{e}(\mathbf{k})= \begin{bmatrix}
        0 & f(\mathbf{k}) \\
        f^*(\mathbf{k}) & 0
    \end{bmatrix} .
\end{equation}
When the the hopping integral $t_0 = -2.7$ eV between the NN $2p_z$ orbitals is considered \cite{graphene_tb}, the Hamiltonian matrix element is $f(\mathbf{k}) = t_0 (1+e^{i{\mathbf{k}} \cdot {\mathbf{a}}_1}+e^{i{\mathbf{k}} \cdot {\mathbf{a}}_2})$. 
Solving the Hamiltonian matrix in Eq. \eqref{eq:graphene_h} yields the electronic band structure of graphene. For a given crystal momentum $\mathbf{k}$, the energy eigenvalue for the band indexed by $n$ is $\varepsilon_{n\mathbf{k}}^0$, and its corresponding Bloch eigenstate is 
\begin{equation}
\varphi_{n\mathbf{k}}^0 =  c_{n,A\mathbf{k}}^0 \varphi_{A\mathbf{k}}^0 + c_{n,B\mathbf{k}}^0 \varphi_{B\mathbf{k}}^0,
\end{equation}
where $n=\{v,c\}$ represents the valence ($v$) and conduction ($c$) bands. Precisely for the crystal momenta $\pm \mathbf{K}$ at the BZ corners [upright inset of Fig. \ref{fig:momentum}(a)], the valence and conduction bands are degenerate, resulting in the characteristic gapless Dirac cones around the $\pm \mathbf{K}$ corners of the BZ. 

\begin{figure}[!b]
    \centering
    \includegraphics[width=1.0\linewidth]{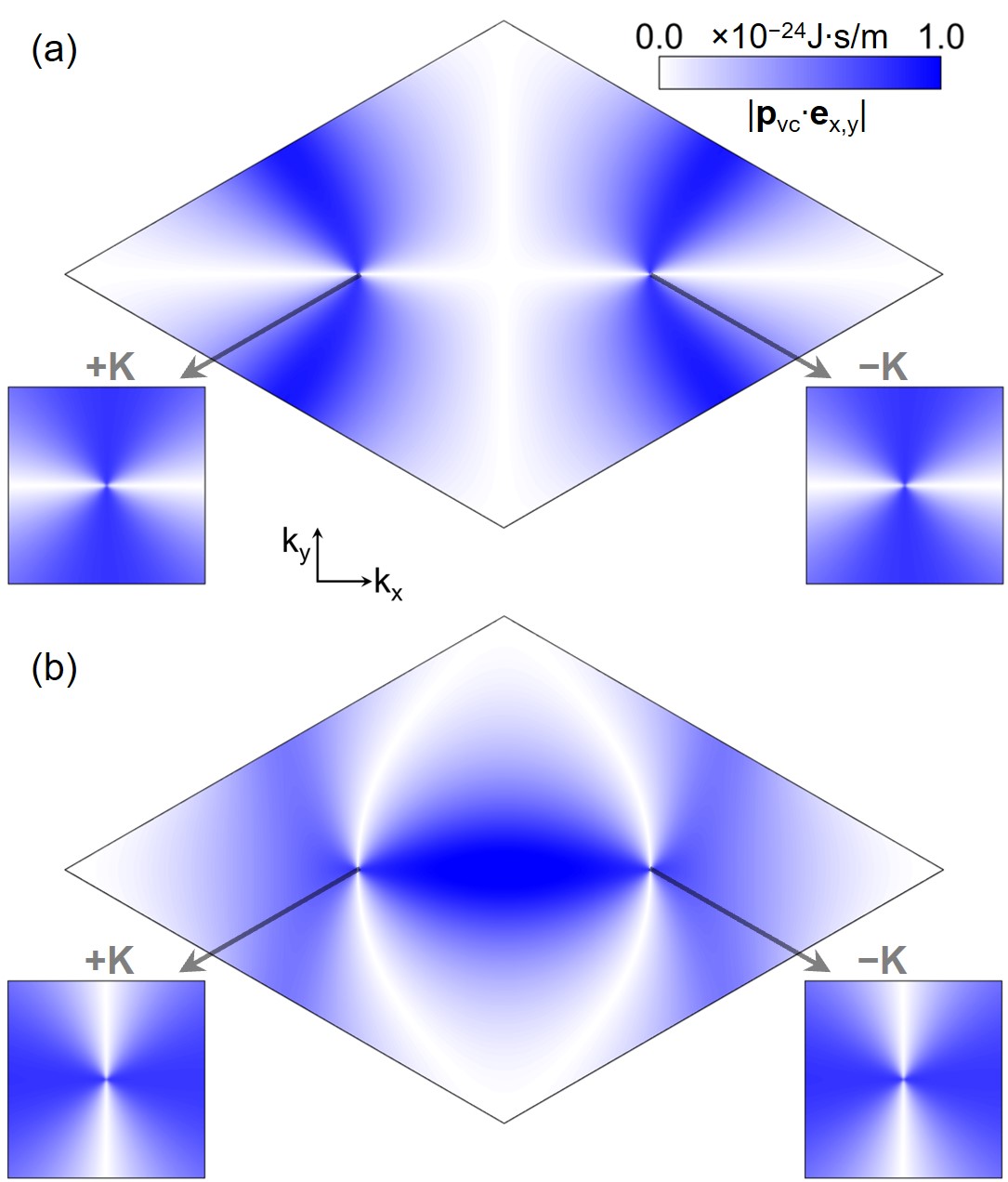}
    \caption{Amplitude of the projection of transition momentum element $\mathbf{p}_{vc\mathbf{k}}$ along $x$ (a) and $y$ directions (b) in the full BZ of graphene. The four square insets show the transition momentum element for the Bloch states at the $\pm \mathbf{K}$ valleys. }
    \label{fig:pvc}
\end{figure}

Next, we derive the matrix elements of the momentum operator, $\hat{\mathbf{p}}=-i\hbar\nabla_{\mathbf{r}}$ with the reduced Planck constant $\hbar$, for the Bloch eigenstates of the two-band graphene model across the full BZ \cite{graphene_momentum_fullTB}. Owing to the spatial periodicity of crystals, the momentum matrix element $ \mathbf{p}_{nn^\prime \mathbf{k}} = -i\hbar \left\langle\varphi_{n\mathbf{k}}^0  |\nabla_\mathbf{r}| \varphi_{n^\prime \mathbf{k}}^0  \right\rangle$ is nonzero only for the two Bloch eigenstates with the same crystal momentum $\mathbf{k}$ (see Appendix \ref{sec:appendix_Fock}). Assuming the value of the gradient operator on the coordinate $\mathbf{r}$ is nonzero only for two NN $2p_z$ orbitals, 
the momentum matrix element is derived as 
\begin{equation}
    \mathbf{p}_{nn^\prime \mathbf{k}} = -i\hbar M_0 \left[  c_{n,A\mathbf{k}}^{0*} c_{n^\prime,B\mathbf{k}}^0  f_{\mathbf{p}} -  c_{n,B\mathbf{k}}^{0*} c_{n^\prime,A\mathbf{k}}^0 f_{\mathbf{p}}^*  \right]
\end{equation}
with $f_{\mathbf{p}} = \frac{\pmb{\delta}_3}{\left|\pmb{\delta}_3\right|}+e^{i \mathbf{k} \cdot \mathbf{a}_1} \cdot \frac{\pmb{\delta}_1}{\left|\pmb{\delta}_1\right|}+e^{i \mathbf{k} \cdot \mathbf{a}_2} \cdot \frac{\pmb{\delta}_2}{\left|\pmb{\delta}_2\right|}$ and $M_0 \approx 5~\text{nm}^{-1}$, where $\pmb{\delta}_1 = \left( \frac{1}{2}, \frac{1}{2\sqrt{3}}  \right) a$, $\pmb{\delta}_2 = \left( -\frac{1}{2}, \frac{1}{2\sqrt{3}}  \right) a$, and $\pmb{\delta}_3 = \left( 0, -\frac{1}{\sqrt{3}}  \right) a$ are the vectors between NN lattice sites. 
Around the crystal momenta $\pm \mathbf{K}$, the electron momentum, described by $\mathbf{p}_{vv\mathbf{k}}$ and $\mathbf{p}_{cc\mathbf{k}}$, shows that the Dirac electrons in graphene has Fermi velocity $v_\text{F} \sim 0.87 \times 10^6$ m/s. 
Precisely for the $\pm{\mathbf{K}}$, the electron momentum exhibits singular behavior due to state degeneracy of the Dirac points, as shown in Fig. \ref{fig:momentum}, where the electron momentum for the $\pm{\mathbf{K}}$ appear as source or drain. Also, the transition momentum elements $\mathbf{p}_{vc\mathbf{k}}$ and $\mathbf{p}_{cv\mathbf{k}}$ exhibit the similar singular behaviors for the states with $\pm{\mathbf{K}}$, as shown in Fig. \ref{fig:pvc}.

At low energy, the Hamiltonian matrix of graphene in Eq. \eqref{eq:graphene_h} can be simplified into an effective Dirac model
\begin{equation}
    h_\text{e}(\mathbf{k})=\hbar v_{\text{F}} \begin{bmatrix}
        0 & k_x + ik_y \\
        k_x - ik_y & 0
    \end{bmatrix},
\end{equation}
where the $\mathbf{k} = (k_x,k_y)$ here in the low-energy model is defined with respect to the $\pm \mathbf{K}$ in the full-BZ model. The Dirac bands with a degenerate point at $\mathbf{k} = (0,0)$ have the perfect linear dispersion, for both the valence band $\varepsilon_{v\mathbf{k}}^0 = -\hbar v_{\text{F}} \sqrt{k_x^2 + k_y^2}$ and conduction band $\varepsilon_{c\mathbf{k}}^0 = -\varepsilon_{v\mathbf{k}}^0$. The momentum operator for Bloch eigenstates can be approximated by $\hat{\mathbf{p}} = \frac{m}{\hbar}\nabla_{\mathbf{k}}\hat{h}_\text{e}$, where the gradient operator is on the coordinate $\mathbf{k}$ of crystal momentum \cite{tight_binding_p_k}. The momentum matrix elements from the low-energy Dirac model are calculated as
\begin{equation}
\label{eq:graphene_momentum}
\begin{aligned}
& \mathbf{p}_{vv\mathbf{k}} = \frac{-mv_{\text{F}}}{\sqrt{k_x^2+k_y^2}}(k_x\mathbf{e}_x+k_y\mathbf{e}_y),~~  \mathbf{p}_{cc\mathbf{k}} = -\mathbf{p}_{vv\mathbf{k}} \\
& \mathbf{p}_{vc\mathbf{k}} = \frac{-imv_{\text{F}}}{\sqrt{k_x^2+k_y^2}}(k_y\mathbf{e}_x-k_x\mathbf{e}_y),~~ \mathbf{p}_{cv\mathbf{k}} = \mathbf{p}_{vc\mathbf{k}}^*
\end{aligned}.
\end{equation} 
Same to that in Fig. \ref{fig:momentum} and \ref{fig:pvc} from the numerical calculations in the full-BZ model, the analytical formula in Eq. \eqref{eq:graphene_momentum} for the low-energy model also indicates a singular behavior of momentum matrix elements, where the electron momentum for states with crystal momenta $\pm \mathbf{K}$ corresponds to source or drain, and the transition momentum elements correspond to vortex. 
With the perfect linear dispersion in the low-energy Dirac model, the momentum matrix elements satisfy $|\mathbf{p}_{vv\mathbf{k}}| = |\mathbf{p}_{cc\mathbf{k}}| =  |\mathbf{p}_{vc\mathbf{k}}|$. Due to the nonlinear component of the Dirac bands in the full-BZ model, the elements satisfy $|\mathbf{p}_{vv\mathbf{k}}| = |\mathbf{p}_{cc\mathbf{k}}| \neq  |\mathbf{p}_{vc\mathbf{k}}|$. In this work, the SI units are employed for all quantities, where the unit of momentum matrix elements is $ \text{J} \cdot \text{s} / \text{m} $, and the unit of mode amplitude $A_0$ is $ \frac{\text{kg} \cdot \text{m}}{\text{C} \cdot \text{s}} $.

\section{QED Fock matrix of graphene coupled with a cavity photon}
\label{sec:appendix_grapheneFock}
The charge neutral graphene in the ground state has total momentum $\mathbf{P} = 0$, thus the QED Fock matrix has the contribution from the cavity-mediated local operator and the induced nonlocal exchange operator with self interaction as in Eq. (\ref{eq:matrixlocal}) and (\ref{eq:matrixexchange}), respectively. For each crystal momentum $\mathbf{k}$, the elements of the $2 \times 2$ QED Fock matrix $F_{\{v,c\}\mathbf{k}}$ in the basis set $\{ \varphi_{v \mathbf{k}}^0 , \varphi_{c \mathbf{k}}^0  \}$ are
\begin{equation}
\label{eq:graphene_hf_matrix}
\begin{aligned}
    & F_{vv\mathbf{k}} =   
 \varepsilon_{v\mathbf{k}}^0 -  \zeta \frac{ {A}_0^{2}}{{\omega}} \big[ \langle \varphi_{v \mathbf{k}}^{0}  |  \hat{\Pi}_{\text{l}} | \varphi_{v \mathbf{k}}^{0}  \rangle  - \langle \varphi_{v \mathbf{k}}^{0}  \varphi_{v \mathbf{k}}  |  \hat{\Pi}_{\text{nl}} | \varphi_{v \mathbf{k}}   \varphi_{v \mathbf{k}}^{0}   \rangle \big] \\
    &    F_{vc\mathbf{k}} =    -  \zeta \frac{ {A}_0^{2}}{{\omega}} \big[ \langle \varphi_{v \mathbf{k}}^{0}  |  \hat{\Pi}_{\text{l}} | \varphi_{c \mathbf{k}}^{0}  \rangle  - \langle \varphi_{v \mathbf{k}}^{0}  \varphi_{v \mathbf{k}}  |  \hat{\Pi}_{\text{nl}} | \varphi_{v \mathbf{k}}   \varphi_{c \mathbf{k}}^{0}   \rangle \big] \\
        &  F_{cv\mathbf{k}} = F_{vc\mathbf{k}} ^* \\
    &    F_{cc\mathbf{k}} =  \varepsilon_{c\mathbf{k}}^0 -  \zeta \frac{ {A}_0^{2}}{{\omega}} \big[ \langle \varphi_{c \mathbf{k}}^{0}  |  \hat{\Pi}_{\text{l}} | \varphi_{c \mathbf{k}}^{0}  \rangle  - \langle \varphi_{c \mathbf{k}}^{0}  \varphi_{v \mathbf{k}}  |  \hat{\Pi}_{\text{nl}} | \varphi_{v \mathbf{k}}   \varphi_{c \mathbf{k}}^{0}    \rangle \big] 
\end{aligned}.     
\end{equation}
By iteratively solving the QED Fock matrix separately for each crystal momentum $\mathbf{k}$ (see Appendix \ref{sec:appendix_Fock} for the iteration scheme), we obtain the cavity-renormalized electronic band structure of graphene, along with the HF orbital wavefunctions, including the occupied states $\varphi_{v \mathbf{k}}$ and the unoccupied states $\varphi_{c \mathbf{k}}$. To ensure that the ground state of graphene in a cavity does not favor $\mathbf{P} \neq 0$, we also construct and iteratively solve the QED Fock matrix without enforcing the condition $\mathbf{P} = 0$. The results obtained are identical to those computed using Eq. \eqref{eq:graphene_hf_matrix}, confirming the zero total momentum in the considered electron-photon coupled system.
Please note, in Eq. \eqref{eq:graphene_hf_matrix}, $\omega$, $A_0$ and $\mathbf{e}$ are the frequency, amplitude and polarization of dressed effective photon modes, where tilde symbol (\textasciitilde) in the notation is omitted for brevity.

\begin{figure}[!t]
    \centering
    \includegraphics[width=1.0\linewidth]{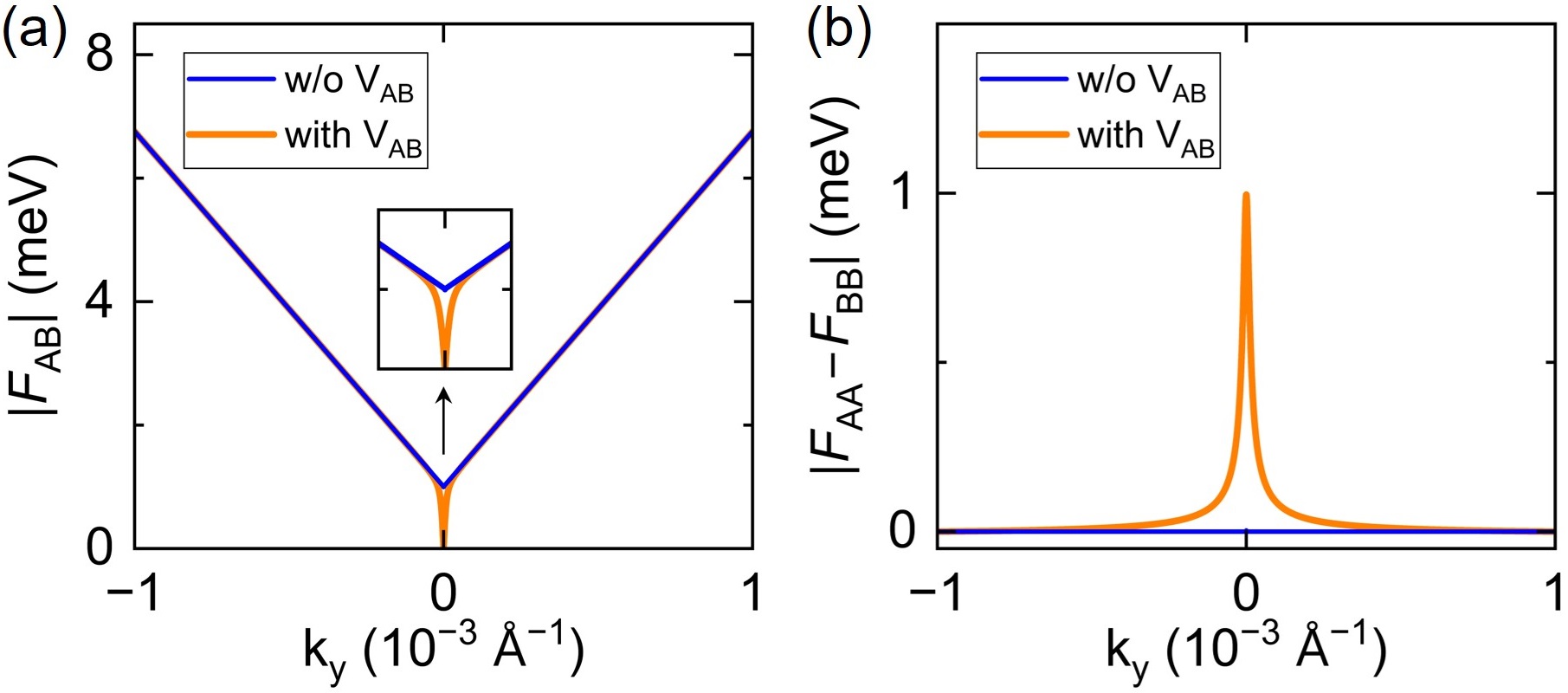}
    \caption{Elements of QED Fock matrix $F_{\{A,B\}\mathbf{k}}$. The frequency, amplitude and polarization of photon mode is the same to that used in Fig. \ref{fig:linear}(a-f). (a) The amplitude of off-diagonal elements, $|F_{AB}|$, for the valence states across the plane $k_x = +K_x$ for $k_y \in +K_y + [-1,1]$~$10^{-3}$\text{\AA}$^{-1}$. Blue (Orange) lines are for graphene with $V_{AB} = 0$ ($\pm 2 \times 10^{-5} t_0$). (b) The amplitude of the difference of diagonal elements $|F_{AA}-F_{BB}|$.}
    \label{fig:fAB_AA_BB}
\end{figure}

As expressed in Eq. \eqref{eq:trans_FAB}, the QED Fock matrix ${F}_{\{v,c\}\mathbf{k}}$ can transformed into ${F}_{\{A,B\}\mathbf{k}}$ in the basis set $\{ \varphi_{A \mathbf{k}}^0 , \varphi_{B \mathbf{k}}^0 \}$. Hopping integrals $t_{AB}$ between $A\text{-}B$ sites contribute to off-diagonal elements $F_{AB}$ and $F_{BA}=F^*_{AB}$, and the hopping integrals $t_{AA}$ ($t_{BB}$) between $A\text{-}A$ ($B\text{-}B$) sites  contribute to diagonal elements $F_{AA}$ ($F_{BB}$). With $V_{AB}=0$, a linearly polarized mode changes $|F_{AB}|$, while does not change $|F_{AA}-F_{BB}|$, as shown in Fig. \ref{fig:fAB_AA_BB}. This indicates the gap opening in Fig. \ref{fig:linear} is contributed from cavity-modified $t_{AB}$, where a long-range spatial anisotropy is induced.   
In comparison, with $V_{AB} \neq 0$, a linearly polarized mode changes both $|F_{AB}|$ and $|F_{AA}-F_{BB}|$ (Fig. \ref{fig:fAB_AA_BB}), showing that the increase in the gap is contributed by the global band change, i.e. across k-points, due to $t_{AB}$, and the local band change around $+\mathbf{K}$ arises from cavity-enhanced difference of $t_{AA}$ and $t_{BB}$. 
By decreasing $V_{AB}$ to zero, the local change from sublattice difference would gradually disappear, and the Dirac gap remains the result of the long-range anisotropic $t_{AB}$ hoppings only. 

\begin{figure*}[!t]
    \centering
    \includegraphics[width=1.0\linewidth]{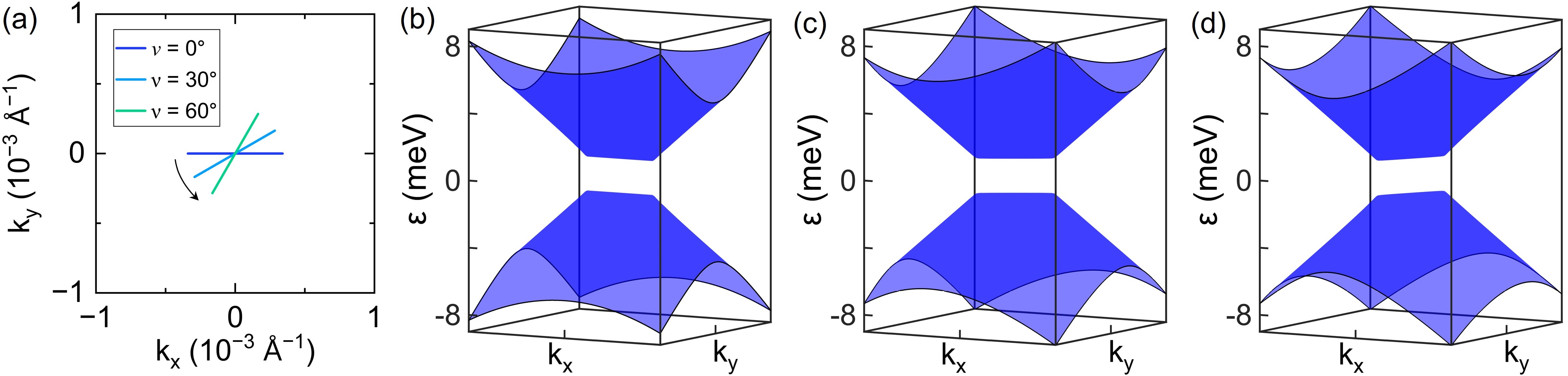}
    \caption{Dirac states in graphene coupled with a linearly polarized photon mode with $\hbar\omega=0.3~\text{eV}$, $A_0 = 2 \times 10^{-8} \frac{\text{kg} \cdot \text{m}}{\text{C} \cdot \text{s}}$, and $\mathbf{e} = (\cos{\nu}) \mathbf{e}_x + (\sin{\nu}) \mathbf{e}_y$. 
    (a) The position of the singular flat-line dispersion in the two-dimensional reciprocal zone $\{k_x , k_y\} \in [-1,1]$~$10^{-3}$\text{\AA}$^{-1}$ centered at $+\mathbf{K}=(k_x,k_y)=(0,0)$. 
    (b,c,d) HF bands, in the $\{k_x , k_y\} \in [-1,1]$~$10^{-3}$\text{\AA}$^{-1}$, induced by the mode with (b) $\nu = 0^\circ$ [same to Fig. \ref{fig:linear}(b)], (c) $\nu = 30^\circ$, and $\nu = 60^\circ$. }
    \label{fig:a3}
\end{figure*}

\section{Calculation of cavity-modified electron density in graphene}
\label{sec:appendix_graphenedensity}
After the solutions of the QED-HF iterations have converged, the cavity-renormalized Bloch wavefunctions of the valence states in graphene are obtained as $\varphi_{v\mathbf{k}} = c_{v} \varphi_{v\mathbf{k}}^0 + c_{c} \varphi_{c\mathbf{k}}^0 $ in the basis set $\{ \varphi_{v\mathbf{k}}^0, \varphi_{c\mathbf{k}}^0 \}$. By a  transformation to the basis set $\{ \varphi_{A\mathbf{k}}^0, \varphi_{B\mathbf{k}}^0 \}$ of the $AB$-site Bloch functions, the wavefunction is expressed as $\varphi_{v\mathbf{k}} = c_{A} \varphi_{A\mathbf{k}}^0 + c_{B} \varphi_{B\mathbf{k}}^0 = \frac{1}{\sqrt{N_\text{cell}}} \sum_\mathbf{R}^{N_\text{cell}} e^{i \mathbf{k} \cdot \mathbf{R}}[c_{A} \varphi_{2p_z}(\mathbf{r}-\mathbf{r}_A)+c_{B} \varphi_{2p_z}(\mathbf{r}-\mathbf{r}_B)] $. The coordinates $\mathbf{r}_A$ and $\mathbf{r}_B$ of lattice sites are shown in Appendix \ref{sec:appendix_grapheneTB}. Then, the electron density is calculated as $\rho{(\mathbf{r})} = \sum_{\mathbf{k}} \rho_{\mathbf{k}} (\mathbf{r})$ with $\rho_\mathbf{k}(\mathbf{r})=|\varphi_{v\mathbf{k}}|^2$, and the cavity-induced density variation is $\Delta \rho (\mathbf{r}) = \rho (\mathbf{r}) - \rho^0 (\mathbf{r})$ with the density $\rho^0 (\mathbf{r}) = \sum_{\mathbf{k}} |\varphi_{v\mathbf{k}}^0|^2$ of graphene outside a cavity.

In the calculation, the $2p_z$ atomic orbital, centered at coordinate origin, is defined as $\varphi_{2p_z} (\mathbf{r}) = \sqrt{\frac{1}{\pi \cdot 2^5} \cdot \frac{Z^3}{a_0^3}} \cdot \frac{Z \cdot z}{a_0} \cdot e^{-\frac{Zr}{2a_0}}$, where $\mathbf{r} = (x,y,z)$, $r = \sqrt{x^2 + y^2 + z^2}$, $Z=3.25$, and $a_0 = 0.529 $~\AA. The $2p_z$ orbital reaches its maximum amplitude at the plane $z_0 = \frac{2a_0}{Z} = 0.33 $~\AA. The crystal momentum $\mathbf{k}$ is sampled using a Monkhorst-Pack grid in reciprocal space.

\section{Evolution of Dirac band anisotropy with polarization direction of a linearly polarized photon}
\label{sec:appendix_rotation}
As shown in Fig. \ref{fig:a3}, with a linearly polarized photon mode, when the polarization direction rotates on the $xy$ plane in real space, the direction of the singular flat line in the QED-HF bands rotates within the $k_x k_y$ plane in reciprocal space, while the size of the induced Dirac gap does not change. The anisotropy of the cavity-renormalized Dirac cones agrees with the anisotropy of the interacting system in real space, where the polarization angle with respect to the $x$ direction in real space is the same to the angle of single-flat-line direction with respect to the $k_x$ direction.

\section{Analytical QED-HF solution for graphene coupled with linearly polarized photons}
\label{sec:appendix_analyticalHF}
With the momentum matrix elements in Eq. \eqref{eq:graphene_momentum} for the effective low-energy Dirac model, the QED Fock matrix
\begin{equation}
\label{eq:graphene_hf_matrix_effective}
    \begin{bmatrix}
        F_{vv\mathbf{k}} & F_{vc\mathbf{k}} \\
        F_{cv\mathbf{k}} & F_{cc\mathbf{k}}
    \end{bmatrix}
    \begin{bmatrix}
        c_{v}  \\
        c_{c} 
    \end{bmatrix}
    =
    \varepsilon
    \begin{bmatrix}
        c_{v}  \\
        c_{c} 
    \end{bmatrix}
\end{equation}
can be analytically constructed for each crystal momentum $\mathbf{k}$ by following Eq. \eqref{eq:graphene_hf_matrix}. The solutions of the matrix for the two crystal momentum paths $k_x = 0$ and $k_y = 0$, passing through Dirac point at $\mathbf{k} = (0,0)$, are analytically derived in the following. 

For a linearly polarized photon with $\mathbf{e} = \mathbf{e}_x$, on the path $k_x = 0$, the energy eigenvalue of the lower valence and upper conduction band is $\varepsilon_v = -\hbar v_\text{F} |k_y| - \frac{\Delta}{2}$ and $\varepsilon_c = \hbar v_\text{F} |k_y| + \frac{\Delta}{2}$, respectively, where the band gap $\Delta = \xi \frac{A_0^2}{\omega} $ with $\xi = 2 \zeta  m^2 v_{\text{F}}^2$ separates the valence and conduction Dirac bands. This indicates that the bands on the path $k_x = 0$ are not smooth when passing through $\mathbf{k} = (0,0)$, showing singularity as discussed in Sec. \ref{sec:a_linearly_polarized_photon}. On the path $k_y = 0$, the Eq. \eqref{eq:graphene_hf_matrix_effective} does not have solutions for $k_x \in [-\frac{L}{2},\frac{L}{2}]$ with $L = \chi \frac{A_0^2}{\omega} $ and $\chi = \frac{2 \zeta  m^2 v_{\text{F}}}{\hbar }$, showing the singularity of the flat-line band with the length $L$ in reciprocal space, as discussed in Sec. \ref{sec:a_linearly_polarized_photon}; the energy eigenvalue for $k_x \notin [-\frac{L}{2},\frac{L}{2}]$ is $\varepsilon_v = -\hbar v_\text{F} |k_x| + \frac{\Delta}{2}$ and $\varepsilon_c = \hbar v_\text{F} |k_x| - \frac{\Delta}{2}$. 

For two isotropic linearly polarized photons with $\mathbf{e} = \mathbf{e}_x$ and $\mathbf{e} = \mathbf{e}_y$, the eigenvalue and eigenstates from the Eq. \eqref{eq:graphene_hf_matrix_effective} keep the same as that without electron-photon interaction, which relies on the equality relation of momentum matrix elements $|\mathbf{p}_{vv\mathbf{k}}| = |\mathbf{p}_{cc\mathbf{k}}| =  |\mathbf{p}_{vc\mathbf{k}}|$. In contrast, with the inequality relation $|\mathbf{p}_{vv\mathbf{k}}| = |\mathbf{p}_{cc\mathbf{k}}| \neq  |\mathbf{p}_{vc\mathbf{k}}|$ from the nonlinear component of Dirac dispersions in the full-BZ model, the Dirac states remain gapless (as that from the low-energy Dirac model here), while the Dirac Fermi velocity is renormalized. This indicates that the cavity-induced variation of Fermi velocity, shown in Sec. \ref{sec:two_linearly_polarized_photons_isotropic}, comes from the inequality relation of momentum matrix elements. Because the inequality is quite small, the renormalization of the Fermi velocity is quite weak at $\pm \mathbf{K}$ valleys, as shown in Fig. \ref{fig:2linear}, which is shown by the significantly weaker states mixing around the $+\mathbf{K}$ valley in Fig. \ref{fig:2linear}(c) than that of a circularly and linearly polarized case in Fig. \ref{fig:circular}(c) and \ref{fig:linear}(c). All the analytical solutions using low-energy effective Dirac model here agree with that in main text using the full-BZ tight-binding model of graphene. 

In addition, in the two-dimensional $\mathbf{k}$ space, we did the numerical photon-free QED-HF calculations using the low-energy effective Dirac model, where the results also agree with the numerical results in main text using the full-BZ tight-binding model. 

\bibliography{main}
\end{document}